\documentclass[aps,prd,10pt,showpacs,preprintnumbers,nofootinbib,superscriptaddress,twocolumn]{revtex4-1}

\usepackage{amsmath,amssymb,amsfonts,amsthm}
\usepackage{bm,bbm}
\usepackage{graphicx}
\usepackage{esint}
\usepackage{color}
\usepackage[breaklinks]{hyperref}
\usepackage{multirow}

\newcommand{\be}{\begin{equation}}
\newcommand{\ee}{\end{equation}}
\newcommand{\ba}{\begin{eqnarray}}
\newcommand{\ea}{\end{eqnarray}}
\def\bs{\begin{subequations}}
\def\es{\end{subequations}}
\def\a{\alpha}

\def\de{\delta}
\def\g{\gamma}
\def\la{\lambda}

\def\e{\epsilon}

\def\om{\omega}
\def\G{\Gamma}

\def\s{\sigma}
\def\vr{\varrho}
\def\vp{\varphi}
\def\N{\nabla}

\def\cD{{\cal D}}

\def\cF{{\cal F}}

\def\cK{{\cal K}}
\def\cL{{\cal L}}
\def\cM{{\cal M}}

\def\cT{{\cal T}}
\def\cV{{\cal V}}
\def\bE{\mathbbm{e}}

\def\ds{d_{\rm S}}
\def\dh{d_{\rm H}}

\def\p{\partial}

\def\B{\Box}
\newcommand{\Eq}[1]{(\ref{#1})}
\def\com{\color{magenta}}
\def\cob{\color{blue}}

\newcommand{\oarX}[1]{\href{http://arxiv.org/abs/#1}{{\ttfamily\com #1}}}
\newcommand{\arX}[1]{\href{http://arxiv.org/abs/#1}{{\ttfamily\com arXiv:#1}}}
\newcommand{\doin}[5]{\href{http://dx.doi.org/#1}{\cob #2 {\bf #3}, #4 (#5)}}
\newcommand{\ndoin}[5]{\href{#1}{\cob #2 {\bf #3}, #4 (#5)}}
\newcommand{\doij}[5]{\href{http://dx.doi.org/#1}{\cob #2 #3 (#5) #4}}
\newcommand{\tia}[1]{}
\newcommand{\boxd}[1]{\boxed{\phantom{\Biggl(}#1\phantom{\Biggl)}}}

\def\rme{e}
\def\rmd{d}
\def\rmi{i}

\begin{document}

\title{Symmetries and propagator in multifractional scalar field theory}

\author{Gianluca Calcagni}
\email{calcagni@iem.cfmac.csic.es}
\affiliation{Instituto de Estructura de la Materia, CSIC, Serrano 121, 28006 Madrid, Spain}
\author{Giuseppe Nardelli}
\email{nardelli@dmf.unicatt.it}
\affiliation{Dipartimento di Matematica e Fisica, Universit\`a Cattolica, via Musei 41, 25121 Brescia, Italy}
\affiliation{INFN Gruppo Collegato di Trento, Universit\`a di Trento, 38100 Povo, Trento, Italy}

\begin{abstract}
The symmetries of a scalar field theory in multifractional spacetimes are analyzed. The free theory realizes the Poincar\'e algebra, and the associated symmetries are modifications of ordinary translations and Lorentz transformations. In the interacting case, the Poincar\'e algebra is broken by interaction terms. The Feynman propagator of the scalar field is computed and found to possess the usual mass poles. As a consequence of these findings, the mass of a particle is a well-defined concept at all scales, and a perturbative quantum theory can be constructed.
\end{abstract}

\date{October 8, 2012}


\preprint{\doin{10.1103/PhysRevD.87.085008}{Phys.\ Rev.\ D}{87}{085008}{2013} [\arX{1210.2754}]}

\maketitle


\section{Introduction}

The ultraviolet (UV) completeness of a quantum field theory depends on the geometry and topology of the spacetime it lives in. Swapping Minkowski spacetime with something more exotic can lead to severely different properties of Feynman diagrams, and, in particular, UV divergences may find a cure. Inspired by the observation, done in standard dimensional regularization, that the renormalization group of a quantum field theory changes with the dimensionality of spacetime \cite{Wil73}, one may wonder what happens on nonconventional geometries with noninteger dimension. In an action formulation, this amounts to replacing the Lebesgue measure in position space with a generic (and possibly very ``irregular'') measure:
\be\label{fractal}
\rmd^Dx\to\rmd\vr(x)\,.
\ee
Interest in these models is further justified, because the sought-for UV improvement should hold also for quantum gravity, where effective geometry changes with the scale in modern approaches (a phenomenon known as dimensional reduction or dimensional flow \cite{tHo93,Car09,fra1,Car10}).\footnote{In the realm of quantum gravity, various are the examples where the effective spacetime dimensionality (whatever ``effective spacetime'' means in each particular approach) changes with the scale: causal dynamical triangulations (where the spectral dimension $\ds$ goes to 2 in the UV) \cite{AJL4,BeH} and the cousin model of random multigraphs \cite{GWZ1,GWZ2}, asymptotic safety (again, $\ds=2$ in the UV) \cite{LaR5,ReS11}, loop quantum gravity and spin foams (where $0\lesssim\ds\lesssim 2$ in the UV) \cite{Mod08,CaM,MPM,COT}, Ho\v{r}ava--Lifshitz gravity ($\ds=2$ in the UV) \cite{Hor2,Hor3,SVW1}, noncommutative geometry at both the fundamental and effective level \cite{Con06,CCM,Ben08,ACOS,AA}, nonlocal superrenormalizable quantum gravity \cite{Mod11,BGKM,AMM,Mod12}, and other approaches \cite{MoN,SSN,Mur12,MuN}. Although there is the perception that dimensional flow and fractal geometry are tightly related to the UV finiteness of all these candidates, the nature of such a relation greatly varies from case to case and is still to be fully understood. Controlling dimensional flow and UV finiteness in a field-theoretical context, eventually drawing a lesson from that and attempting to apply it to quantum gravity, partly motivates the present work.} Early attempts to formulate field theories on fractal spacetime date back to the 1970s and 1980s \cite{Sti77,Svo87,Ey89a,Ey89b}, but since then little progress has been made due to severe technical challenges stemming from Eq.\ \Eq{fractal} \cite{fra1,Gol08,fra2,fra3}. In fact, not only Poincar\'e symmetries are broken, but also the continuous texture of spacetime.

Recently, it was proposed to tackle the problem in a continuum framework where the measure in \Eq{fractal} is a factorizable measure found in fractional calculus \cite{ACOS,frc1,frc2,frc3,frc4,frc5,fra7} (overviews and a review are in Refs.\ \cite{fra4,fra6,AIP}). Fractional measures are known to describe random fractals and to approximate measures on deterministic fractals \cite{RYS,YRZ,RYZLM,Yu99,QL,RQLW,RLWQ,NLM}; they have anomalous scaling properties directly leading to an effective dimension of spacetime different from its integer topological dimension $D$. Fractional measures not only allow one to define a momentum space and a transform between this and position space, but they are also naturally prone to generalization to multiscale geometries (in particular, Refs.\ \cite{frc4,fra6,fra7} are devoted to dimensional flow in fractional spacetimes and other approaches). Progress in and applications of the model are ongoing but still at an early stage. For instance, power-counting renormalizability is easy to prove for a scalar field theory with integer-order d'Alembertian and an arbitrary perturbative potential \cite{frc2}, and applications to particle-physics phenomenology are promising \cite{HX}, but the quantum theory was not analyzed in detail. In particular, fractal geometry modifies symmetries, Noether currents, and unitarity of a scalar field theory \cite{fra2}, but what happens exactly for a fractional measure is not clear. The study of quantum mechanics on fractional spacetimes suggests that unitarity be violated in a controllable way \cite{frc5}, but little else is known about quantum fractional systems. Because of breaking of Poincar\'e invariance, it is not even obvious that the concepts of mass and spin for a particle field still make sense. Only after answering these and other elementary questions could one attempt to extract phenomenological consequences and continue the investigation of renormalization properties.

The purpose of this paper is to discuss the symmetries and quantization of a scalar field theory on multifractional Minkowski spacetime and to find the propagator. Fractional spacetime and its generalization to generic factorizable measures are reviewed in Secs.\ \ref{rev} and \ref{revfm}. In Sec.\ \ref{mfst}, we construct the multiscale version of fractional Minkowski spacetime by using a new factorizable \emph{Ansatz}, differing from the definition in Refs.\ \cite{fra4,frc2,frc4} but sharing the same properties, which we shall analyze. The great advantage yielded by a factorizable version of the multifractional measure is to permit a well-defined momentum space and ``Fourier'' transform and, hence, an automatic extension of the fixed-scale results to a geometry with dimensional flow.

We begin the analysis of the scalar field theory with a discussion of the symmetries at the classical level (Sec.\ \ref{csft}). Somewhat against intuition, the free theory does admit a symmetry with generators satisfying the \emph{ordinary} Poincar\'e algebra. Consequently, under the assumption of the adiabatic switching of the interaction, it is still possible to define mass and spin of a particle as the (asymptotic) eigenvalues of the Casimir operators of a suitable representation of the  Poincar\'e group. The generators of this algebra, however, do not correspond to usual Poincar\'e transformations, which are ``deformed'' by nontrivial measure factors. The classical interacting theory is worked out in Sec.\ \ref{roma}. The energy-momentum tensor and the other Noether currents are not conserved, and the algebra in the infinite-dimensional representation is broken. Noether currents for the general case of a Lebesgue--Stieltjes measure were obtained in Ref.\ \cite{fra2}. However, here we point out that only measures admit a clean treatment, while that of Ref.\ \cite{fra2} should be regarded as qualitative. In Sec.\ \ref{propa}, we quantize the free field and show how creation and annihilation operators also exist on fractional spacetimes. The Feynman propagator in both position and momentum space is then found. All these results are valid, in particular, for factorizable multifractional as well as log-oscillating measures. Section \ref{disc} is devoted to discussion.


\section{Multiscale Minkowski spacetime}\label{revmf}

We begin with fractional Minkowski spacetime $\cM^D_\a$ with fixed dimensionality, later moving to the context of the multiscale generalization, which is straightforward at the level of position space.


\subsection{Spacetime: Review}\label{rev}

$\cM^D_\a$ is defined via a $D$-dimensional embedding charted by coordinates $x^\mu$, $\mu=0,1,\dots, D-1$. In this case, the embedding is $M^D$, ordinary Minkowski spacetime with metric $\eta_{\mu\nu}={\rm diag}(-,+,\cdots,+)$. Next, one defines a nontrivial measure of the form
\bs\label{amea}\ba
&&\rmd\vr_\a(x)=\rmd^Dx\,v_\a(x)\,,\\
&& v_\a(x)=\prod_\mu v_\a(x^\mu):=\prod_\mu \frac{|x^\mu|^{\a_\mu-1}}{\Gamma(\a_\mu)}\,,
\ea\es
where $\G$ is the gamma function and $\a_\mu$ are $D$ real parameters (``fractional charges'') in the range $0<\a_\mu\leq 1$. In the simplest ``isotropic'' case, $\a_\mu=\a$ are all equal. Complex fractional charges are also possible, and their rich effects on the geometry are discussed in Refs.\ \cite{ACOS,fra4,frc2}.

The geometric inequivalence between free fractional and integer theories is first illustrated by the Hausdorff dimension. The latter can be inferred from the scaling law of the measure:
\be\label{scala}
\vr_\a(\la x)=\la^{\dh}\vr_\a(x)\,,\qquad \dh=\sum_\mu\a_\mu\,,
\ee
where it is implicit that the dimensionality of coordinates $x$ is fixed; in particular, $[x^\mu]=-1$ in momentum units, for all $\mu$. In the isotropic case, $\dh=D\a$. The Hausdorff dimension can be determined also by the scaling of the volume of a $D$-ball with radius $R$:
\be\label{vr}
\cV^{(D)}(R) = \int_{D\text{-ball}} \rmd \vr_\a(x)\propto R^{\dh}\,.
\ee
As noted in Sec.\ 3.5 of Ref.\ \cite{frc1}, one can call $q^\mu=\vr_\a(x^\mu)$, obtain the Lebesgue measure $d^Dq$, and calculate the volume of the ball in $q$ coordinates; the final result, with the original length units, is clearly the same. The reason is that reparametrizations do not modify the momentum-space structure of the theory, which is defined so that coordinates $x$ have length dimension (therefore, $q$ coordinates have anomalous scaling by definition). This guarantees, in general, that fractional measures represent nontrivial geometries. The scaling laws \Eq{scala} and \Eq{vr} can be found in phenomenological models of fractal or porous media \cite{Tar3,Tar4b,Tar7} and are typical of measures on mathematical sets with fractal geometry \cite{Fal03} and of Lebesgue--Stieltjes measures \cite{fra1,Pod02}, of which fractional measures are a special case. As mentioned in the introduction, in fact, fractional measures do represent fractal geometries, either under certain approximations or exactly \cite{RLWQ,NLM,tat95}.

The geometry of fractional spacetimes is further characterized by the spectral dimension $\ds$ which, together with $\dh$, determines the scaling of correlation functions. In that case, the existence of a probability density function $P$ governing a diffusion process on these spacetimes, together with the self-similar scaling property of $P$, shows that a test particle diffuses anomalously in fractional spacetimes \cite{frc1,frc4}, as it happens on fractals and various complex media \cite{Sok12}. These properties are well known in anomalous transport systems, where they find several applications \cite{Sok12,MeK,Zas3,MeK2}. On the other hand, diffusion is normal in ordinary spacetimes, and $\dh=\ds=D$ in the absence of curvature.

Minkowski fractional spacetime can be equipped with generalizations of the Laplace--Beltrami (or d'Alembertian) operator $\B=-\p_t^2+\N^2_x$. In general, one can devise self-adjoint operators ``$\p^{2\g}$'' of fractional order $2\g$ \cite{frc4}. Here we choose the second-order operator \cite{frc3}
\be\label{ka}
\cK_\a=\eta^{\mu\nu}\cD_\mu\cD_\nu\,,\quad \cD_\mu:=\frac{1}{\sqrt{v_\a(x)}}\,\p_\mu\left[\sqrt{v_\a(x)}\,\,\cdot\,\right],
\ee
which is self-adjoint with respect to the natural scalar product. By adopting $\cK_\a$, integrations by parts are considerably simpler, and there are no nonlocal effects. Also, the propagator will turn out to have only poles (i.e., particle modes), not branch cuts \cite{frc2}, where one would see quasiparticle modes of unclear physical interpretation (related to nonlocality). With a second-order operator, the theory can have a UV critical point associated with a two-dimensional spacetime and power-counting renormalizability; this is a good signal that the field theory has improved UV properties, which should be verified by explicit renormalization techniques. Power-counting renormalizability in general fails for theories with fractional Laplacians, although renormalizability in those cases is not excluded, either. 

The natural exterior derivative on spaces endowed with the Laplace--Beltrami operator \Eq{ka} is suggested by the form of the weighted derivatives $\cD_\mu$:
\be\label{bard}
\rmd_v := \frac{1}{\sqrt{v_\a}}\, \rmd \left(\sqrt{v_\a}\,\cdot\,\right)\,,
\ee
where $\rmd$ is the ordinary differential defined as
\be\label{damu}
\rmd := \rmd x^\mu\p_\mu\,.
\ee
Multiplying the members of Eq.\ \Eq{damu} by $1/\sqrt{v_\a}$ to the left and $\sqrt{v_\a}$ to the right, we obtain
\be\label{damu2}
\rmd_v = \rmd x^\mu \cD_\mu\,.
\ee
Notice that the right-hand side features the ordinary differential of the coordinates, so that one can keep using the ordinary differential for all purposes. 
 Equations \Eq{damu} and \Eq{damu2} give a differential structure obviously different from the one of fractional spaces with noninteger derivatives \cite{frc1}, where the fractional exterior derivative \cite{Tar12,CSN1} is employed.

Momentum space has the same structure of $\cM_\a^D$ except that its measure may be different, $v_{\a'}(k)$ possibly with $\a'\neq \a$. Noting that the eigenfunctions of $\cK_\a$ can be normalized as (here $k^2:=k_\mu k^\mu=-k_0^2+\sum_{i=1}^{D-1}k_i^2$)
\bs\label{E}\ba
&&\bE_v(k,x) = \frac{1}{\sqrt{v_{\a'}(k)v_\a(x)}}\,\frac{\rme^{\rmi k\cdot x}}{(2\pi)^{\frac{D}{2}}}\,,\\ &&\cK_\a\bE_v(k,x)=-k^2\bE_v(k,x)\,,
\ea\es
we can find an invertible momentum transform \cite{frc3}:
\bs\label{fmt}\ba
\tilde f(k) &:=& \int_{-\infty}^{+\infty}\rmd\vr_\a(x)\,f(x)\,\bE_v^*(k,x)=:F_{v}[f(x)]\,,\label{fo1}\\
f(x) &=& \int_{-\infty}^{+\infty}\rmd\vr_{\a'}(k)\,\tilde f(k)\,\bE_v(k,x)\,.\label{fo2}
\ea\es
If $\a'=\a$, $F_v$ is an automorphism. Unitarity of the transform is guaranteed by the integral representation of the fractional delta distribution,
\bs\label{delfra}\ba
\de_\a(x,x') &:=&\frac{\de(x-x')}{\sqrt{v_\a(x)v_\a(x')}}\nonumber\\
&=& \int\rmd\vr_{\a'}(k)\,\bE_v^*(k,x)\bE_v(k,x')\,,\\
\de_{\a'}(k,k')&:=&\frac{\de(k-k')}{\sqrt{v_{\a'}(k)v_{\a'}(k')}}\,,
\ea\es
generalizing the Dirac distribution and the concept of pointwise source in fractional spacetimes.

Extended localized sources can be packed through weighted Gaussian distributions ($D=1$ example)
\be
H_\s(x) = \frac{\rme^{-\frac{x^2}{8 \s}}}{(4 \pi \s)^{1/4} \sqrt{v_\a (x)}}\,,
\ee
which are $L^2_\vr ({\mathbb R})$ [i.e., square integrable on the real axes with fractional measure $\vr (x)$].
Note that $H_\s(x)\in C^\infty_{\cal D}$; i.e.,  it is differentiable infinitely many times with respect to the weighted derivative $\cal D$.

 The momentum transform \Eq{fmt} applied to $H_\s(x)$ induces a momentum Gaussian distribution
\be
\tilde H_{\s'}(k) = \left(\frac{4 \s}{\pi}\right)^{1/4} \frac{\rme^{- 2 \s k^2}}{\sqrt{v_{\a'}(k)}} \equiv \frac{\rme^{-\frac{k^2}{8 \s'}}}{(4 \pi \s')^{1/4} \sqrt{v_{\a'} (k)}}\,,
\ee
with  a variance $\s'$  which is  dual to that of $H_\s(x)$, namely, $4 \s'=1/(4 \s)$. The  momentum transform \Eq{fmt} preserves the Parseval identity, and $\|H_\s(x)\|^2_{\vr(x)} = \|\tilde H_{\s'}(k)\|^2_{\vr'(k)}$,
as one can easily check. 

The Fourier transform of $H_\s (x)$ defining the momentum Gaussian distribution can be interpreted as a fractional integral of an exponential generating function of the type $h(x; \s,k)= \rme^{-x^2/(8 \s) - \rmi k x}$. As such, in the spirit of Wilson \cite{Wil73}, any integral of any (suitably well behaved)  function  can be obtained by using differentiation with respect to $k$ and/or  $\s$. With this procedure, Feynman diagrams can be defined as functionals (e.g., Ref.\ \cite{Svo87}) on a space of rapidly decreasing functions  
\ba
{\cal S}_{\vr} ({\mathbb R})&=&\left\{ \phi(x)\in C^\infty_{\cal D}\vphantom{\sup_\mathbb{R}}\ ,\right.\nonumber\\
&&\quad\left. \sup_\mathbb{R} \bigl| x^m \cD^n \phi(x) \bigr|<\infty , \ \forall\, m,n\in\mathbb{N}\right\}.
\ea
Starting from $H_\s(x)$, a complete  basis of ${\cal S}_{\vr} ({\mathbb R})$ can be easily obtained, either by multiplication of $H_\s(x)$ by suitable polynomials (Hermite) or by an arbitrary number of weighted derivatives acting on $H_\s(x)$. Since 
$ {\cal S}_{\vr} ({\mathbb R})$ is dense in $L^2_\vr ({\mathbb R})$ with respect to the $L^2_\vr ({\mathbb R})$ topology, the same set of functions serves also as a complete basis for $L^2_\vr ({\mathbb R})$.


\subsection{Generalization to factorizable measures}\label{revfm}

Unless we want to compute the Hausdorff or spectral dimension or solve dynamical equations, we can be less specific about the form of the measure weight $v_\a$ \cite{frc3,frc5} and ask only for two properties:
\begin{enumerate}
\item[(1)] The measure weight must be \emph{factorizable} in the coordinates:
\ba
\rmd \vr(x)&:=&\rmd^Dx\,v(x)=\rmd t\, v_0(t)\,\rmd{\bf x}\, v({\bf x})\nonumber\\
					 &:=&\rmd t\, v_0(t)\,\prod_{i=1}^{D-1}\rmd x^i\,v_i(x^i)\,,\label{genf}
\ea
where the $D$ functions $v_\mu$ can be all different. To avoid confusion later with spacetime indices, we denote the spatial part with no index as $v({\bf x})$.
\item[(2)] The measure weight must be positive semidefinite (as before, the subscript $\mu$ in the weight is not a vector index):
\be\label{v>0}
v_\mu=v_\mu(x^\mu)\geq 0\,.
\ee
\end{enumerate}
A third condition, that $v_0=v_0(|t|)$ is even in time, is not necessary here, although it guarantees a unitary limit for the $S$-matrix in quantum mechanics \cite{frc5}.

With general measure, we denote the Laplacian $\cK_\a$ simply as $\cK_v$ and the delta distribution $\de_\a$ as $\de_v$. Momentum-space measure $\rmd\tau(k)=\rmd^Dk\,w(k)$ can be functionally different from the one in position space, but to lift the burden from the notation we shall use the same symbols, $\rmd\vr(k)=\rmd^Dk\,v(k)$. We still call these spacetimes ``fractional,'' in the broader meaning of ``factorizable.''


\subsection{Multifractional and log-oscillating spacetimes}\label{mfst}

Fractional spaces have fixed dimensionality, and, unless $\a=1$, they do not describe the geometry we observe. To get a physically viable setting, it is necessary to let the effective dimension change with the scale. Mimicking the definition of self-similar measures of multifractal geometry \cite{Hut81,Har01}, one considers linear superpositions of fractional measures on a discrete set of charges $\a_n$ \cite{frc2,frc4}. In one dimension, we have
\be\label{addi}
v(x)=\sum_{n=1}^N h_n v_{\a_n}(x)\,,
\ee
where $h_n>0$ are couplings with dimension (length)$^{1-\a_n}$ and $N$ is finite. They define $N$ regimes where the Hausdorff dimension is approximately constant and given by $\dh\sim\a_n$. The simplest case of dimensional flow is realized by \emph{binomial} measures, where $N=2$ and there is only one fundamental scale.

In Refs.\ \cite{fra4,frc2,frc3,fra6,frc4}, the following generalization to $D$ topological dimensions was adopted:
\be\label{muf1}
v_{\rm diag}(x):=\sum_n h_n \left[\prod_\mu v_{\a_n}(x^\mu)\right]\,.
\ee
The sum is performed over $D$-dimensional measures: The idea is that a multifractal in a given $D$-dimensional embedding be realized by taking ``snapshots'' at different scales. Each snapshot corresponds to a fractal with fixed Hausdorff dimension [i.e., a fractional space with measure \Eq{amea}]. Unfortunately, the measure \Eq{muf1} is the sum of factorizable measures, so it is not factorizable and hinders the construction of an invertible momentum transform.

Here we change the way snapshots are taken, and, rather than choosing the multiscale generalization of the $D$-dimensional measure, we generalize to $D$ dimensions the one-dimensional multiscale measure. Therefore, we consider the product of $D$ multifractals in one-dimensional embeddings:
\be\label{muf2}
v_*(x):=\prod_\mu v_*(x^\mu):=\prod_\mu\left[\sum_n g_n v_{\a_n}(x^\mu)\right]\,,
\ee
where $g_n=h_n^{1/D}$ (with engineering dimension $[g_n]=\a_n-1$) so that Eq.\ \Eq{muf1} represents all the diagonal terms of $v_*$. This prescription is now factorized and all the momentum-space formalism, the form of Laplacians, of quantum operators, and so on, will be valid for $v=v_*$. Equation \Eq{muf2} with $g_n$ replaced by some $g_n^{(\mu)}$ was briefly mentioned in Ref.\ \cite{frc3}, but there it was not welcomed on the ground that the dimensionality along each direction would flow independently, while one might expect that a given $D$-dimensional configuration evolves as a whole throughout the probed scales. However, anisotropic dimensional flow is conceivable in its own right \emph{a priori}, and, if one wants an isotropic change of geometry, it is sufficient to take the couplings $g_n^{(\mu)}=g_n$ for all $\mu$. Thus, there is no conceptual problem with \Eq{muf2} and its anisotropic counterpart. 

\begin{figure}
\centering
\includegraphics[width=7cm]{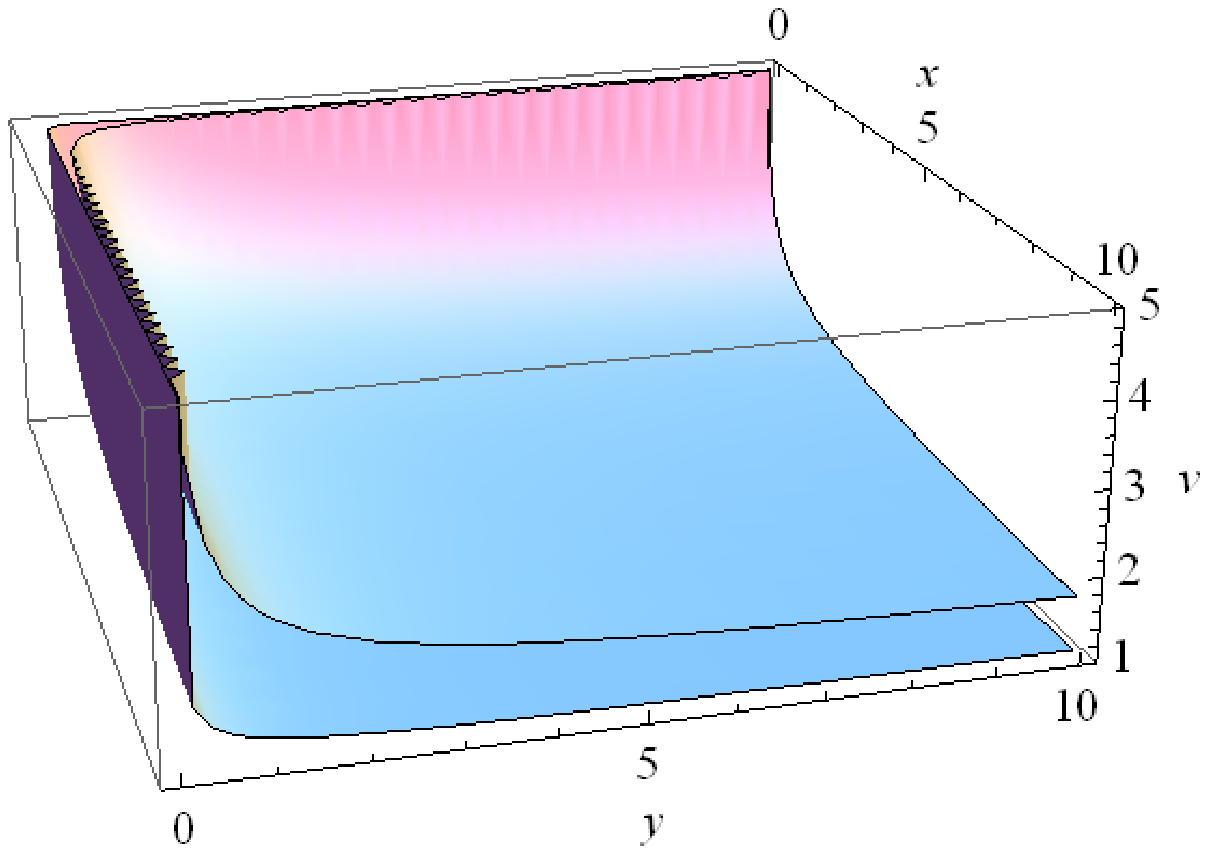}
\includegraphics[width=7cm]{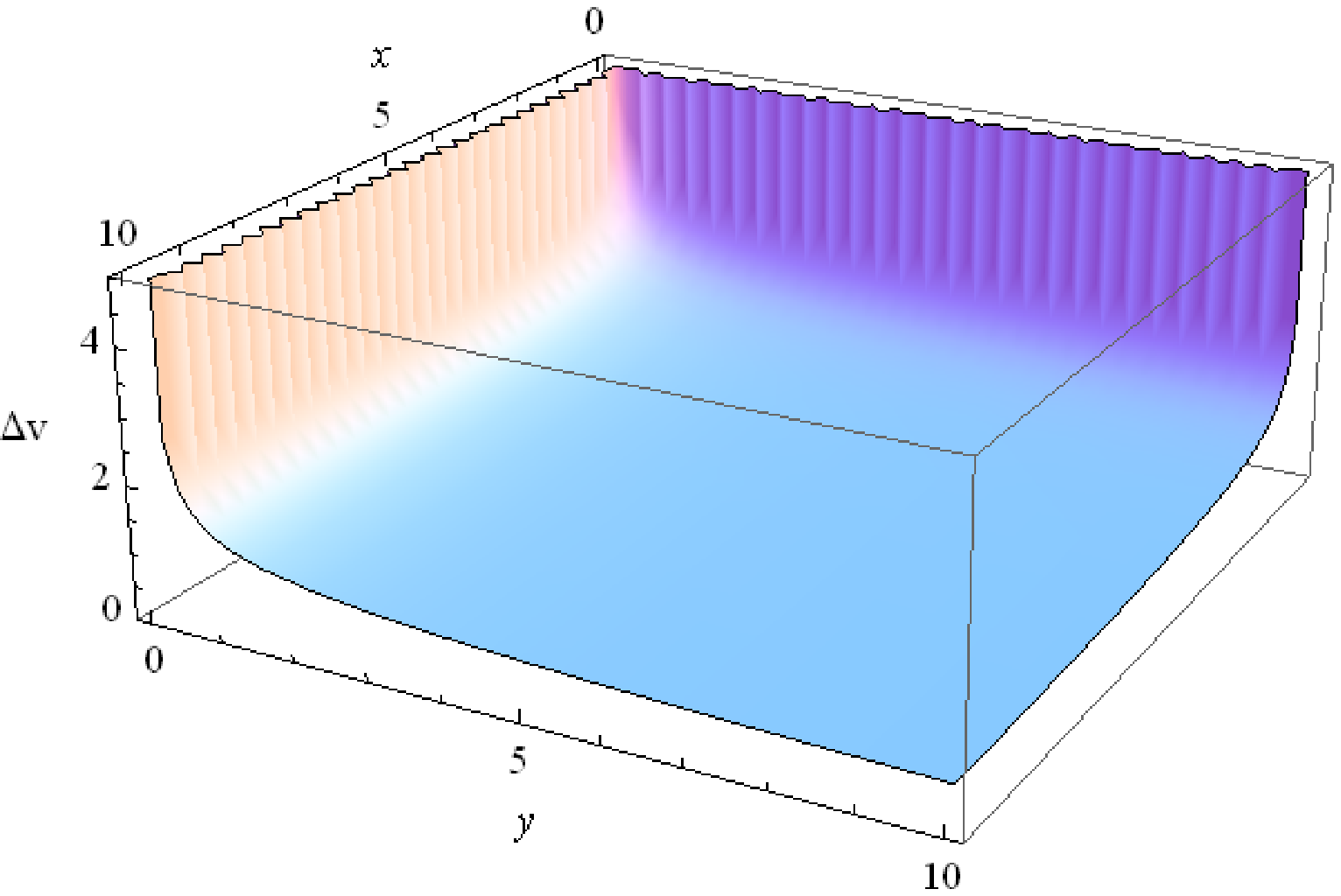}
\caption{\label{fig1} Top panel: The nonfactorizable measure $v_{\rm diag}$ [Eq.\ \Eq{muf1}, bottom surface] and the factorizable one $v_*$ [Eq.\ \Eq{muf2}, top surface] for $D=2$, $\a_1=1/2$ and $\a_2=1$. Bottom panel: The difference $\Delta v=v_*-v_{\rm diag}$. Gamma factors have been omitted.}
\end{figure}

One might still wonder about the effect of the ``off-diagonal'' terms $\Delta v=v_*-v_{\rm diag}$, but they are subdominant along each direction with respect to the diagonal part $v_{\rm diag}$. They change only the slope of the measure weight; this slope can be adjusted by tuning the magnitude of the couplings. Figure \ref{fig1} shows a two-dimensional binomial ($n=1,2$) example with $\a_1=1/2$ and $\a_2=1$. Here $v_*=(1+|x|^{-1/2})(1+|y|^{-1/2})$ and $v_{\rm diag}=1+|xy|^{-1/2}$. As one can see, from the point of view of measure behaviour, little changes qualitatively.

One can also check that \Eq{muf2} gives the correct dimensional flow by computing the Hausdorff dimension $\dh$ and the spectral dimension $\ds$. The calculation of $\dh$ and $\ds$ in the factorizable multifractional case is very similar to the one in Refs.\ \cite{frc2,AIP} and Ref.\ \cite{frc4}, respectively. For the reader's convenience, here we recall the result for the Hausdorff dimension for a binomial measure and in the isotropic case. For simplicity, we consider one topological dimension, $D=1$, and a configuration where the geometry reduces to the standard one at large scales. The binomial measure is then $v_*(x)=1+|x/\ell_*|^{\a_*-1}/\Gamma(\a_*)$, where $0<\a_*< 1$ and $\ell_*$ is some fixed length, a scale characteristic of the geometry which precisely discriminates between ``small'' and ``large'' distances. The volume of a ball can be calculated as before [Eq.\ \Eq{vr}], yielding (the prefactors are the volumes of unit balls)
\be
\cV^{(1)}(R)= 2R\left[1+\frac{1}{\Gamma(1+\a_*)}\left(\frac{R}{\ell_*}\right)^{\a_*-1}\right]\,.
\ee
The result is easily generalizable to $D$ dimensions, with the care that for the factorizable measure there are also unimportant cross terms. Thus, we can identify two geometric regimes. In the isotropic case, these are
\bs\ba
R\ll\ell_*:&\qquad& \cV^{(D)}\sim R^{D\a_*}\,,\\
R\gg\ell_*:&\qquad& \cV^{(D)}\sim R^{D}\,.
\ea\es
Given a probing scale $\ell$, the Hausdorff dimension runs from $\dh(\ell\ll\ell_*)\sim D\a_*$ in the ultraviolet to $\dh(\ell\gg\ell_*)\sim D$ in the infrared.

We conclude with yet another generalization, when the fractional charges in Eq.\ \Eq{muf2} are promoted to complex parameters, $\a\to\a+\rmi\om$ \cite{ACOS,fra4,frc2}. Measures of this type can be rendered real by combining conjugate pairs of weights, thus generating an oscillatory pattern periodic in $\ln (|x|/\ell_\infty)$, where $\ell_\infty$ is a fundamental scale which can be identified with the Planck length \cite{ACOS}. In one dimension, these measures are of the form
\ba
v_{\a,\om}(x)&:=& v_\a(x)\left\{1+A_{\a,\om}\cos\left[\om\ln\left(\frac{|x|}{\ell_\infty}\right)\right]\right.\nonumber\\
&&\left.+B_{\a,\om}\sin\left[\om\ln\left(\frac{|x|}{\ell_\infty}\right)\right]\right\}\,,
\ea
where $A$ and $B$ are real. Log-oscillating geometries occur frequently in chaotic systems \cite{Sor98} and are characteristic of self-similar fractals \cite{LvF}. In fact, complex fractional calculus is a better approximation of deterministic fractals than real-order calculus \cite{NLM}, and, in this sense, it may be regarded as a more complete description of anomalous geometries. Moreover, complex measures associated with deterministic fractals can have concrete physical applications in, for instance, quantum systems and statistical mechanics \cite{Akk1,Akk2,Akk12}.

Log-oscillating measures are endowed with discrete symmetries. Because of this and the natural presence of a fundamental scale, when applied to spacetime itself they give rise to a rich hierarchy of scales and geometry regimes with intriguing physical interpretations \cite{ACOS,frc2}. In the present work, we mention only that we can let oscillating measures fall in the category of positive factorizable measures \Eq{genf} and \Eq{v>0} provided $A_{\a,\om}^2+B_{\a,\om}^2\leq 1$. Then, $v_{\a,\om}\geq 0$, and one can include in the forthcoming discussion also
\be\label{co2}
\bar v(x):=\prod_\mu \bar v(x^\mu):=\prod_\mu\left[\sum_{n,l} g_{\a_n,\om_l} v_{\a_n,\om_l}(x^\mu)\right]\,,
\ee
where $g_{\a_n,\om_l}>0$.

The results of Ref.\ \cite{frc5} on quantum mechanics in anomalous spacetimes are thus valid also for multifractional and log-oscillating measures when selecting \Eq{muf2} and \Eq{co2}.


\section{Classical scalar field theory}\label{csft}


\subsection{Fractional and integer pictures}\label{twopi}

A scalar field with potential $V$ living in factorizable (and, in particular, multifractional) Minkowski spacetime is governed by the action
\ba
S&=&\int_{-\infty}^{+\infty}\rmd\vr(x)\left[\frac12\phi\cK_v\phi-V(\phi)\right]\nonumber\\ &=&\int_{-\infty}^{+\infty}\rmd\vr(x)\left[-\frac12\cD_\mu\phi\cD^\mu\phi-V(\phi)\right]\,,\label{S}
\ea
where in the last step we took advantage of the weights in $\cD$ to integrate by parts and obtain a quadratic form. In the integration by parts, boundary terms evaluated at $x=\pm\infty$ can be omitted by requiring suitable asymptotic behaviour of the fields, whereas possible  boundary terms at $x^\mu=0$ vanish under the assumption of continuity of $\p_\mu [\sqrt{v(x)} \phi(x)]$ at the origin.

Notice that, by defining
\be\label{pshi}
\vp(x):=\sqrt{v(x)}\,\phi(x)\,,
\ee
for a power-law potential $V(\phi)\propto\phi^n$ the action \Eq{S} becomes
\bs\label{Spsi} \ba
S&=&\int_{-\infty}^{+\infty}\rmd^Dx\,\bar\cL\,,\\
\bar\cL &=&-\frac12\p_\mu\vp\p^\mu\vp-[v(x)]^{1-\frac{n}2} V(\vp)\,,
\ea\es
which is an action in ordinary spacetime but with a nonautonomous (i.e., explicitly coordinate-dependent) potential. The quadratic potential ($n=2$, mass term) does not have any nonautonomous factor, and in the free (quadratic) Lagrangian the fractional measure can be completely reabsorbed in a field redefinition. We say the theory \Eq{S} expressed in fractional spacetime to be in the ``fractional'' or ``scalar density picture,'' while the one in the field variable $\vp$ is in the ``integer'' or ``scalar picture.''

To justify these names, we notice that $\phi$ is a scalar with respect to the symmetries of ordinary spacetime, but it is a \emph{scalar density} of weight $-1/2$ from the point of view of fractional spacetime (so $\vp$ is a genuine scalar).
 
Consider first a generic field $f(x)$ in ordinary spacetime and an infinitesimal coordinate transformation $x\to x'=x+\de x$. As is well known, for the field $f$ two distinct transformations can be defined. The first is the functional transformation $f(x)\to f'(x) := f(x)+  \de f(x)$, which is the field transformation evaluated at the same point. This is the field variation used to recover the Euler--Lagrange equations of motion. The complete transformation $\de_0 f (x)$ should also take into account the corresponding  coordinate transformation, namely, $f(x)\to f'(x') := f(x)+\de_0 f(x)$. This transformation typically defines the ``type'' of field under the specific transformation $\delta x$. For example, in Minkowski spacetime all the fields are usually taken to be scalars under translations, whereas under Lorentz transformations they can be scalars, spinors, vectors, and so on.
Clearly, for infinitesimal transformations the following relation holds:
\be
\label{dede}
\de_0 f(x) = \delta f(x) + \delta x^\mu \p_\mu f(x)\,.
\ee


If $f=\vp$ is a scalar field under translations, $\de_0 \vp=0$. In ordinary spacetime, ${x'}^\mu=x^\mu+\de x^\mu=x^\mu-\e^\mu$ is a translation ($\e$ is constant) and in this case the interpretation of Eq.\ \Eq{dede} is that the functional variation $\de \vp(x)$ should compensate the original coordinate transformation to guarantee $\de_0 \vp=0$. The (infinite-dimensional) Hermitian generator of translations in Minkowski spacetime is the operator $\hat p_\mu = - i \p_\mu $. Accordingly, the unitary infinite-dimensional representation of translations $\e^\mu$  is given by $\bar U_\e:= \rme^{i \e^\mu \hat p_\mu}= \rme^{\e^\mu \p_\mu}$, and the finite functional variation reads
  \be
  \label{fifuvp}
   \vp'(x) = U_\e \vp(x) = e^{\e^\mu \p_\mu}\vp(x) = \vp(x+\e)\,.
  \ee
 As expected, the above transformation is precisely the one needed to compensate the coordinate transformation $x\to x-\e$, so that $\vp$ is indeed a scalar under translations: The infinitesimal version of \Eq{fifuvp} is
 $\delta \vp(x) = \e^\mu \p_\mu \vp(x)$, and since  $\delta x^\mu = -\e^\mu $, \Eq{dede} gives
\be\label{devp}
\de_0 \vp(x) =\e^\mu \p_\mu \vp(x)-\e^\mu \p_\mu \vp(x)= 0\,.
\ee
This is the definition of scalar field under translations.
  
In factorizable spacetimes with measure weight $v(x)$, the operator $\hat p_\mu = - i \p_\mu$ is not self-adjoint with respect to the natural scalar product. The self-adjoint derivative operator is rather [see Eq.\ \Eq{ka}]
\be 
\label{phatdef}
\hat P_\mu:=-\rmi\cD_\mu\,.
\ee
This is the generator  of ``fractional translations,'' so that their unitary infinite-dimensional representation is \cite{frc5}
\be 
U_\e:=\rme^{\rmi\e^\mu \hat P_\mu}=\frac{1}{\sqrt{v}}\,\bar U_\e\,\sqrt{v}\,.
\ee
The action of a fractional translation on a scalar function $\phi(x)$ can be easily inferred from Eq.\ \Eq{pshi}, leading to 
\ba
\phi'(x)&:=&U_\e\phi(x)=\frac{1}{\sqrt{v(x)}}\,\vp(x+\e)\nonumber\\
&=&\sqrt{\frac{v(x+\e)}{v(x)}}\,\phi(x+\e)\,.\label{Uphi}
\ea
In infinitesimal form, expanding \Eq{Uphi} up to $O(\e)$,
\be \label{deph}
\de\phi=\e^\mu\cD_\mu\phi\,,
\ee
so that
\ba
\de_0\phi&=&\e^\mu\cD_\mu\phi-\e^\mu\p_\mu\phi=\frac12 \e^\mu \frac{\p_\mu v}{v}\,\phi\nonumber\\
&=& -\frac12 \delta  x^\mu  \frac{\p_\mu v}{v}\,\phi\neq 0\,,\label{bardep}
\ea
and, as anticipated, with respect to translations in fractional spacetimes, $\phi$ is a scalar density of weight $-1/2$. Equations \Eq{devp} and \Eq{deph} are mutually consistent, because $\de$ is a field (not coordinate) variation, and $\de \vp=\de(\sqrt{v}\phi)=\sqrt{v}\de\phi$. It also follows that $\de \p_\mu\vp=\p_\mu\de\vp$ and
\be \label{decom}
\de\cD_\mu\phi=\cD_\mu\de\phi\,.
\ee

In the next sections we will employ only the scalar density picture, which is the fundamental one, leaving parallel calculations in the integer picture to the Appendix. 


\subsection{Functional variations and equations of motion}

In fractional spaces, functional derivatives $\de_v$ should be such that
\be 
\frac{\de_v \eta(x)}{\de_v \eta(y)}=\de_v(x,y)=\frac{\de({x}-{y})}{\sqrt{v(x)v(y)}}\,,
\ee 
leading to the relation
\be \label{devr}
\frac{\de_v}{\de_v \eta(x)} =\frac{1}{v(x)}\frac{\de}{\de \eta(x)}
\ee
for any function $\eta(x)$. [If the functional is integrated  only on spatial coordinates, then the prefactor is $1/v({\bf x})$.] 

As an important note of caution, we stress that, although Eq.\ \Eq{devr} is formally (i.e., as a local limit) ill defined at the measure singularities [if any, as $x=0$ in the original fractional case \Eq{amea}] and zeros, both the scalar and fractional formulations of the theory are well defined, because \Eq{devr} is in the sense of distributions and measure factors cancel one another. 

Functional Taylor expansions are independent from the picture adopted (fractional or integer): For any functional $A$,
\ba  
A[f+ \eta] &=& A[f] +  \sum_{n=1}^\infty \frac{1}{n!} \int d^Dy_1\cdots d^Dy_n\nonumber\\
&&\qquad\times  \frac{\delta^n  A[f]}{\delta f (y_1) \cdots \delta f (y_n)} \, \eta(y_1) \cdots \eta(y_n)\nonumber\\
&=&  A[f] +  \sum_{n=1}^\infty \frac{1}{n!} \int d\vr(y_1)\cdots d\vr (y_n)\nonumber\\
&&\qquad\times \frac{\de_v^n  A[f]}{\de_v f (y_1) \cdots \de_v  f (y_n)} \, \eta(y_1) \cdots \eta(y_n)\,,\nonumber\\ 
\ea
where in the last equation we made use of \Eq{devr}. Consequently,
distinction between ordinary and fractional variation is immaterial for the equations of motion, since $\de_v S=0$ implies $\de S=0$ including at measure singularities, if any. Eventually, using \Eq{decom} one finds
\be\label{eom}
\frac{\p\cL}{\p\phi}-\cD_\mu\frac{\p\cL}{\p(\cD_\mu\phi)}=0\,,
\ee
that is,
\be\label{eomph} 
\boxd{\cD_\mu\cD^\mu\phi-V_{,\phi}(\phi)=0\,.}
\ee


\subsection{Free theory: Poincar\'e symmetry}

Fractional spacetimes break translation invariance, and, consequently, ordinary momentum is not conserved. In spite of this fact, it is possible to define mass and spin for a free particle and a propagator whose poles in momentum space are precisely the masses of the one-particle states. Clearly, consistence with translation noninvariance requires that the momentum operator is not the usual translation generator. On the other hand, mass and spin can be unambiguously defined only for theories where the generators satisfy the Poincar\'e algebra. This is precisely the route we shall follow: (i) Find the infinite-dimensional representations of operators $\hat P$ and $\hat J$  satisfying the Poincar\'e algebra
\bs\label{poin}\ba 
&& [\hat P_\mu,\hat P_\nu]=0\,,\label{PP}\\
&& [\hat P_\mu,\hat J_{\nu\rho}]=\rmi(\eta_{\mu\rho}\hat P_\nu-\eta_{\mu\nu}\hat P_\rho)\,,\label{PJ}\\
&& [\hat J_{\mu\nu},\hat J_{\s\rho}]=\rmi (\eta_{\mu\rho}\hat J_{\nu\sigma}-\eta_{\nu\rho}\hat J_{\mu\sigma}+\eta_{\nu\s}\hat J_{\mu\rho}-\eta_{\mu\s}\hat J_{\nu\rho})\,;\nonumber\\ \label{JJ}
\ea\es
(ii) find a vector space whereupon these operators act; (iii) find the eigenstates of $\hat P^2$ and $\hat W^2$ (where $\hat W^\mu=\e^{\mu\nu\rho\s}\hat P_\nu \hat J_{\rho\s}/2$ is the Pauli--Lubanski pseudovector). For a local relativistic theory, there is the further requirement that the vector space found in (ii) be invariant under representations of $\hat P$ and $\hat J$. 

It turns out that the fractional \emph{free} theory does admit a Poincar\'e algebra \Eq{poin}. The only difference between the standard case and fractional spacetimes is that the operators $\hat P$ and $\hat J$ will not generate ordinary Poincar\'e transformations but something slightly more exotic. In the interacting theory, these deformed Poincar\'e transformations will be further modified, and the Poincar\'e algebra will be broken.

We show that we can identify the generators in \Eq{PP} with the ones defined in \Eq{phatdef}. In fact, it is immediate to verify that the latter commute, so that \Eq{PP} is satisfied. Furthermore, 
in the action \Eq{S}, the kinetic term and a mass term (quadratic potential) are invariant under the action of the transformations \Eq{Uphi}. These are immediate consequences of the transformation properties of $\phi$ described in Sec.\ \ref{twopi}. From \Eq{PP}, $[U_\e,\hat P_\mu]=0$, and hence
\ba
\label{ptra}
\hat P_\mu'\phi'(x)&:=&U_\e(\hat P_\mu\phi)=\hat P_\mu(U_\e\phi)\nonumber \\
&=& -i \frac{1}{\sqrt{v(x)}} \p_\mu \left[ \sqrt{v(x)} \ \sqrt{\frac{v(x+\e)}{v(x)}}\phi(x+\e) \right]\nonumber \\
&=&-i \sqrt{\frac{v(x+\e)}{v(x)}}\ \frac{1}{\sqrt{v(x+\e)}}\nonumber\\
&&\qquad\times\p_\mu \left[ \sqrt{v(x+\e)}\, \phi(x+\e) \right]\nonumber\\
&=& \sqrt{\frac{v(x+\e)}{v(x)}} [\hat P_\mu \phi](x+\e)\,,
\ea
where in the last step we made explicit the coordinate dependence of $\hat P$. Therefore, an action with a kinetic term of the type $\hat P_\mu \phi \hat P^\mu \phi$ is invariant under \Eq{Uphi}:
\ba
&&\int \rmd^Dx\, v(x)\, [\hat{P}_\mu{}'\phi'](x)\,[\hat{P}^\mu{}'\phi'](x)\nonumber\\
&&\quad=\int \rmd^Dx\, v(x+\e)\, [\hat P_\mu\phi](x+\e)\,[\hat P^\mu\phi](x+\e)\nonumber\\
&&\quad=\int \rmd^Dx'\, v(x')\, [\hat P_\mu\phi](x')\,[\hat P^\mu\phi](x')\,.
\ea
A mass term is also invariant, since
\ba 
\int \rmd^Dx\, v(x) {\phi'}^2(x)&=&\int \rmd^Dx\, v(x+\e) \phi^2(x+\e)\nonumber\\
&=&\int \rmd^Dx'\, v(x') \phi^2(x')\,.
\ea
Consequently, the fractional  translation operators $\hat P$ do indeed generate transformations which are symmetries of the free action. Higher-order potentials are not invariant under fractional translations.

Fractional Lorentz transformations are generated by the operator
\be 
\hat J_{\nu\rho}:=x_\nu \hat P_\rho-x_\rho\hat P_\nu= \frac{1}{\sqrt{v}}\,\hat\jmath_{\nu\rho}\,\sqrt{v}\,.
\ee
As before, this operator is not associated with ordinary rotations and boosts, which are generated by $\hat\jmath_{\nu\rho}:=x_\nu \hat p_\rho-x_\rho\hat p_\nu$. From the
commutation relations  between $\hat p$ and $\hat\jmath$, it follows immediately that Eq.\ \Eq{PJ} holds.\footnote{A direct calculation is lengthier. One shows that $\hat P_\mu \hat J_{\nu\rho}\phi=-\rmi \eta_{\mu\nu}\hat P_\rho\phi+x_\nu \hat P_\mu \hat P_\rho\phi$, while $\hat J_{\nu\rho}\hat P_\mu\phi=x_\nu \hat P_\rho \hat P_\mu\phi$. Taking the difference of these expressions and using \Eq{PP}, one obtains Eq.\ \Eq{PJ}.} The same route leads to verification of \Eq{JJ}.

The action of a fractional Lorentz transformation on a field $\phi(x) $ is defined by $\phi'(x)= U_\omega \phi(x)$, where $U_\omega$ is the infinite-dimensional unitary representations of the fractional Lorentz transformations $\hat J$ with parameters $\omega$:
\bs\label{Uomega}\ba
U_\omega &:=& \rme^{-\frac{\rmi}2 \omega^{\mu \nu} \hat J_{\mu \nu}}=\frac{1}{\sqrt{v}}\,\bar U_\omega\,\sqrt{v}\,,\\
\bar U_\omega &:=& \rme^{-\frac{\rmi}2 \omega^{\mu \nu} \hat\jmath_{\mu \nu}}\,.
\ea\es
Equation \Eq{Uomega} with $\hat J$ replaced by the usual $D$-dimensional generators of Lorentz transformations defines the corresponding $D$-dimensional Lorentz matrix $\Lambda^\mu_{\ \nu}\approx\de^\mu_\nu+\om^\mu_{\ \nu}$. It is not difficult to verify that the action \Eq{S} with quadratic potential is indeed invariant under the transformation generated by \Eq{Uomega}. This is a consequence of the counterparts of Eqs.\ \Eq{Uphi} and \Eq{ptra}\footnote{For instance, the first expression can be verified as follows: $U_\om\phi(x)=v^{-1/2}\bar U_\om [v^{1/2}\phi(x)]=v^{-1/2}\bar U_\om \vp(x)=v^{-1/2}\vp(x')$. We omit the direct calculation, which is based on the infinitesimal transformation $U_\omega \approx \mathbbm{1}-(\rmi/2)\omega^{\mu \nu} \hat J_{\mu \nu}$.}:
\bs\ba
U_\om\phi(x) &=& \sqrt{\frac{v(x')}{v(x)}}\,\phi(x')\,,\\\
U_\om[\hat P_\mu\phi(x)]&=&\sqrt{\frac{v(x')}{v(x)}}\,[\hat P_\mu'\phi](x')\,,
\ea\es
where
\be\label{lore}
{x'}^\mu=\Lambda^\mu_{\  \nu} x^\nu
\ee
and $\hat P_\mu'=\Lambda_\mu^{\ \nu }\hat P_\nu$.  Although the Lorentz transformation on the coordinates are the usual $D\times D$ matrices, the free action is invariant under the fractional representation of the Poincar\'e group via the operators $\hat P_\mu$ and $\hat J_{\mu\nu}$. 

Since a scalar field has vanishing spin, the study of Lorentz transformations and related invariants does not add more information on the fields. Classically, all the Noether invariants associated with Lorentz transformations can be written solely in terms of the energy-momentum tensor (the angular momentum is purely orbital). Quantum mechanically, a scalar field generates states that belong to the trivial representation of $\hat W^2$. Consequently, from now on we shall concentrate mainly on the transformations generated by $\hat P$.
 
 

\subsection{Energy-momentum tensor}

The derivation of the energy-momentum tensor highlights the subtlety in the transformation property of the scalar density $\phi$. In the following, we denote
\be
\pi^\mu:=\frac{\p\cL}{\p(\cD_\mu\phi)}=-\cD_\mu\phi\,.
\ee
Consider a coordinate and field transformations $x\to x+\de x$, $\phi\to \phi+\de_0\phi$. Then,
\ba  
\de_0 S &=&\int\left[\de(\rmd^Dx)v\cL+\rmd^Dx\cL\de_0 v+\rmd^Dx\,v\de_0\cL\right]\nonumber\\
&=&\int\rmd^Dx[v\cL\p_\mu\de x^\mu +\cL\p_\mu v\de x^\mu\nonumber\\
&&\qquad+v(\de\cL+\p_\mu\cL\de x^\mu)]\nonumber\\
&=&\int\rmd^Dx\left[\p_\mu(v\cL\de x^\mu)+ v\de\cL\right]\nonumber\\
&\ \stackrel{\Eq{decom}}{=}\ &\int\rmd^Dx\left[\p_\mu(v\cL\de x^\mu)+ v\left(\frac{\p\cL}{\p\phi}\de\phi+\pi^\mu\cD_\mu\de\phi\right)\right]\nonumber\\
&\ \stackrel{\Eq{eom}}{=}\ &\int\rmd^Dx\left[\p_\mu(v\cL\de x^\mu)+ v\left(\cD_\mu\pi^\mu\de\phi+\pi^\mu\cD_\mu\de\phi\right)\right]\nonumber\\
&=&\int\rmd^Dx\,\p_\mu\left(v\cL\de x^\mu+ v\pi^\mu\de\phi\right)\nonumber\\
&=&\int\rmd^Dx\,\p_\mu\left[\left(v\de^\mu_\nu\cL-v\pi^\mu\cD_\nu\phi\right)\de x^\nu\right]\nonumber\\
&=\ &\int\rmd^Dx\,\p_\mu\left(vT^\mu_\nu\de x^\nu\right)\,,
\ea
where in the next-to-last step we used the generalization of Eq.\ \Eq{deph} to an arbitrary coordinate transformation, $\de\phi=-\de x^\nu\cD_\nu\phi$, and we defined
\be\label{Tmunu}
\boxd{T_{\mu\nu}:=\eta_{\mu\nu}\cL+\cD_\mu\phi\cD_\nu\phi\,.}
\ee

In ordinary Minkowski spacetime, the energy-momentum tensor satisfies a continuity equation $\p_\mu \bar T^\mu{}_\nu =0$ by virtue of the fact that translation is a symmetry of the system, $\de_0 S=0$. This does not happen in the fractional case. In fact, for a translation, $\de x^\mu=-\e^\mu$ is constant, and one ends up with
\be
\de_0 S = -\int\rmd^Dx\,\p_\mu\left(vT^\mu_\nu\right)\e^\nu\,.\label{app1}
\ee
The left-hand side does \emph{not} vanish because $S$ is not left invariant by a coordinate transformation due to the potential:
\ba
\de_0 S &=& \int \rmd^D x'\,v(x')\cL'(x')-\int \rmd^D x\,v(x)\cL(x)\nonumber\\
          &=& -\int \rmd^D x\left\{v(x-\e)V'[\phi(x-\e)]-v(x)V[\phi(x)]\right\}\nonumber\\
          &=& -\int \rmd^D x\left[-\e^\nu\p_\nu v V(\phi)+vV_{,\phi}\de_0\phi\right]\nonumber\\
          &\ \stackrel{\Eq{bardep}}{=}\ & -\int \rmd^D x\,\p_\nu v\left(\frac{1}{2}\phi V_{,\phi}-V\right)\e^\nu\,.\label{app2}
\ea

Equating \Eq{app1} to \Eq{app2} and due to the arbitrariness of $\e^\nu$, we obtain 
\be
\label{contphi}
\p_\mu  (v T^\mu{}_\nu) = \p_\nu v\left(\frac{1}{2}\phi V_{,\phi}-V\right)\,.
\ee
Equation \Eq{contphi} is not yet in its final form. In fact, the components of $T^\mu{}_\nu$ are scalar densities with weight $-1$ with respect to translations (they are bilinears of scalar densities of weight $-1/2$). Thus, it follows that the natural derivative acting on them in fractional spaces is
\be
\label{weight1}
\check{\cD}_\mu:=\frac{1}{v}\p_\mu\left(v\,\cdot\,\right)\,.
\ee
Taking this on board, the continuity equation reads
\be\label{contphi2}
\boxd{\check{\cD}_\mu T^\mu{}_\nu =\frac{\p_\nu v}{v}\left(\frac{1}{2}\phi V_{,\phi}-V\right) =: s_\nu(x,\phi)\,.}
\ee
Therefore, $T^\mu{}_\nu$ is not conserved in the usual sense even in the free theory ($n=2$), when the right-hand side of Eq.\ \Eq{cont} vanishes identically but the derivative in the left-hand side still carries a hidden self-source term.

Integrating \Eq{contphi2} on a spatial slice, with appropriate (vanishing) boundary conditions of the fields at spatial infinity, we obtain the conservation equation
\be\label{contphi3}
\boxd{\check{\cD}_t P^\nu = \int \rmd\vr({\bf x})\,s^\nu(x,\phi)\,,}
\ee
where the momentum $D$-vector $P^\nu$ is defined by
\be\label{Pnu}
P^\nu := \int \rmd\vr({\bf x})\, T^{0\nu}\,,
\ee
and $\rmd\vr({\bf x}) = \rmd{\bf x} \, v({\bf x})$. The explicit components read
\ba  
 \label{ham1} H &:=& P^0 = \int\rmd\vr({\bf x})\,\left[\frac12 \pi_\phi^2+\frac{1}{2}\cD_i\phi\cD^i\phi+V(\phi)\right],\\
 \label {Pi} P^i &=& - \int\rmd\vr({\bf x})\, \pi_\phi \cD^i\phi\,,
\ea
where
\be\label{piph} 
\pi_\phi:= \cD_t\phi\,.
\ee

As expected, the fractional momentum is never conserved in time, not even for a quadratic potential [that makes the right-hand side of \Eq{contphi3} vanish]. In particular, for $\nu=0$ the fractional Hamiltonian $H$ is not conserved in time due to dissipation from the fractional measure.


\section{Interacting theory: deformed Poincar\'e algebra}\label{roma}

In this section we show that the time-space translation algebra does \emph{not} close. The same type of steps lead to an analogous conclusion for Lorentz boosts and rotations. Therefore, in the free theory there is an accidental Poincar\'e invariance which is broken in the interacting case.

Equal-time fractional Poisson brackets are defined as\footnote{Since integration is only over spatial variables, the fractional functional derivative has a factor $1/v({\bf x}) $. See the comment below Eq. \Eq{devr}.}
\ba
\{A({\bf x}),B({\bf x'})\}_v&:=& \int \rmd\vr({\bf y})\, \left[\frac{\de_v A({\bf x})}{\de_v\phi({\bf y})}\frac{\de_v B({\bf x'})}{\de_v\pi_\phi({\bf y})}\right.\nonumber\\
&&\qquad\left.-\frac{\de_v A({\bf x})}{\de_v\pi_\phi({\bf y})}\frac{\de_v B({\bf x'})}{\de_v\phi({\bf y})}\right]\,,\label{poia}
\ea
which imply the following  canonical fractional Poisson brackets: 
\bs\label{canopoi}\ba
\{\phi(t,{\bf x}),\pi_\phi(t,{\bf x'})\}_v&=& \de_v ({\bf x,x'})\nonumber\\
&=& \frac{\de({\bf x-x'})}{\sqrt{v({\bf x}) v({\bf x'}) }}\,,\\
\{\phi(t,{\bf x}),\phi(t,{\bf x'})\}_v&=& 0\,,\\
\{\pi_\phi(t,{\bf x}),\pi_\phi(t,{\bf x'})\}_v&=& 0\,.
\ea\es
There follows the relation
\be \label{poipoi}
\{\cdot,\cdot\}_v = v_0(t)\,\{\cdot,\cdot\}\,,
\ee
where it is understood that the brackets in the right- and left-hand side are computed via the conjugate pair $(\phi,\pi_\phi)$ and $(\vp,\pi_\vp)$, respectively, where $\pi_\vp=\dot\vp=\sqrt{v}\pi_\phi$.

Canonical fractional  brackets  allow us to write the equations of motion as Hamilton equations, showing that $H$ is the canonical generator of fractional time translations:
\be\label{d0ev}
\cD_t\phi =\{\phi,H\}_v\,,\qquad \cD_t\pi_\phi =\{\pi_\phi,H\}_v\,.
\ee
The first equation coincides with \Eq{piph}, while the second is the equation of motion \Eq{eomph}. 
A cross-check can be done by using the scalar picture described in the Appendix: Comparing \Eq{hamps} with \Eq{d0ev} reproduces  condition \Eq{poipoi}. Similarly, from Eqs. \Eq{Pi} and \Eq{poia} it immediately follows that 
\be\label{diev}
-\cD^i\phi =\{\phi,P^i\}_v\,,\qquad -\cD^i\pi_\phi =\{\pi_\phi, P^i\}_v\,,
\ee
so that indeed the momentum  $P^\nu$ associated with the Noether current is the generator of weighted derivatives through fractional Poisson brackets:
\be
-\cD^\mu\phi =\{\phi,P^\mu\}_v\,.
\ee
Finally,  we shall verify under which conditions the momentum algebra is satisfied. Clearly, 
$\{H,H\}_v=0$ by antisymmetry. The bracket $\{P^i,P^j\}_v$ can be made to vanish under suitable boundary conditions of the fields at infinity. In fact, it is a boundary term: From Eqs.\ \Eq{poia} and \Eq{Pi}, we get
\begin{widetext}
\ba
\label{pipj}
\{P^i,P^j\}_v&=&\int \rmd \vr({\bf x})\rmd \vr({\bf y})\, \{\pi_\phi (t, {\bf x}) \cD^i \phi(t,{\bf x}),\pi_\phi (t, {\bf y}) \cD^j \phi(t,{\bf y}) \}_v \nonumber \\
&=&\int \rmd \vr({\bf x}) \, \left[ \cD^i \phi(t,{\bf x}) \cD^j \pi_\phi(t,{\bf x}) -\cD^j \phi(t,{\bf x}) \cD^i \pi_\phi(t,{\bf x})\right]\nonumber\\
&=&\int \rmd {\bf x} \, \left[  \p^i\left( \sqrt{v({\bf x})} \phi (t,{\bf x}) \right) \p^j\left( \sqrt{v({\bf x})} \pi_\phi (t,{\bf x}) \right)- \left( i \leftrightarrow j \right)\right] \nonumber\\
&=&\int \rmd {\bf x} \,   \p^i\left[ \sqrt{v({\bf x})} \phi (t,{\bf x}) \p^j\left( \sqrt{v({\bf x})} \pi_\phi (t,{\bf x}) \right)\right ] - \left( i \leftrightarrow j \right)=0\,.
\ea
The mixed bracket in general does not vanish. By using Eqs.\ \Eq{poia}, \Eq{ham1}, \Eq{Pi}, and \Eq{eomph}, it can be written, on shell, as 
\ba
\label{piH}
\{P^i, H \}_v&=&-\int \rmd \vr({\bf x})\rmd \vr({\bf y})\,\left\{\pi_\phi (t, {\bf x}) \cD^i \phi(t,{\bf x}),\frac12 (\pi_\phi (t, {\bf y}))^2 +\frac12  (\cD_j \phi(t,{\bf y}))^2 + V(\phi(t,{\bf y})) \right\}_v \nonumber \\
&=&\int \rmd \vr({\bf x}) \, \left[ \pi_\phi (t,{\bf x}) \cD^i \pi_\phi(t,{\bf x}) +\cD^i \phi(t,{\bf x}) \left( -\cD_j \cD^j \phi(t,{\bf x}) + V_{,\phi}(\phi) \right)\right]\nonumber\\
&=&\int \rmd \vr({\bf x}) \, \left[ \cD_t \phi (t,{\bf x}) \cD^i \pi_\phi(t,{\bf x}) -\cD^i \phi(t,{\bf x}) \cD_0 \pi_\phi (t,{\bf x})\right]\nonumber\\
&=& \left[\p_t+\frac{\dot v_0(t)}{v_0(t)}\right]\left[ -\int \rmd \vr({\bf x})\pi_\phi (t, {\bf x}) \cD^i \phi(t,{\bf x})\right]=\dot P^i+\frac{\dot v}{v}\,P^i=\check{\cD}_t P^i\,.
\ea
\end{widetext}
This last bracket does not vanish in fractional spaces, spoiling the momentum algebra of the general theory.\footnote{On the contrary, in the usual Minkowskian case the right-hand side of \Eq{piH} is $\dot P^i$, which vanishes as $P^i$ is a conserved Noether charge.} This is not surprising, since the theory is not translation invariant.  In such a case, the physical interpretation of the quantum theory could be troublesome. The mass of a particle state is the eigenvalue of the operator $-P_\mu P^\mu$ only if $P^\mu$ satisfies the Poincar\'e algebra, and, consequently, it would be unclear what a theory describes at the quantum level if Poincar\'e symmetry were missing. Fortunately, this is not exactly the case we are considering, and we are in an intermediate position: Perturbation theory can be consistently defined. In a quantum theory, perturbative particle states are generated by creation operators in a spacetime region where the interaction is adiabatically switched off. This means that, to clearly identify (and label) the physical parameters characterizing a particle (such as its mass and spin), it is understood that the particle should be free: There exist asymptotic quantum field operators $\phi_{\rm in}$ and $\phi_{\rm out}$ satisfying \emph{free} equations of motion, and the physical Hilbert spaces is built through repeated action of asymptotic (free) creation operators on the vacuum. ``Free fields''  means fields with quadratic potential and, from Eqs.\ \Eq{piH}, \Eq{contphi2}, and \Eq{contphi3}, it follows that 
\be
\{P^i,H\}_v = \int\rmd {\bf x}\,  \p^i v({\bf x})\left[ \frac12 \phi\, V_{,\phi} (\phi) - V(\phi) \right]\,,
\ee 
which does vanish for $V\propto\phi^2$. The momentum algebra closes for free theories, and asymptotic quantum states can be unambiguously defined.

At this point, the physical interpretation of the system enters into a sharper focus. Having established that fractional integrals approximate certain fractal sets (Sec.\ 4.4 of Ref.\ \cite{frc1}) and following the physical interpretation of fractional integrals given in classical mechanics (Sec.\ 2.6 of Ref.\ \cite{frc1}), we end up with a two-part physical system. Fractional mechanical systems are dissipative \cite{frc5,Pel91,Rie1,Rie2,El05a,El05b}, where the order $\a$ is the fraction of states surviving at a given time. These states can be imagined to be distributed inside a fractal $\cF$, while states located outside $\cF$ are lost. The same picture holds in fractal spaces or spacetimes, in particular in the fractional realization. Here, an observer lives inside fractional Minkowski spacetime $\cM_\a^D$ (with nontrivial measure weight), in turn embedded in the ambient space $M^D$ (the bulk, with ordinary Lebesgue measure), Minkowski space. According to the observer (fractional picture), the integer Poincar\'e algebra is deformed and Noether charges are not conserved; at the quantum level, this corresponds to a loss of unitarity (see also \cite{frc5,Pel91}), parametrized by the fractional charge $\a$. Energy and information are lost in the bulk. From the perspective of the ambient space (integer picture) this is an exchange of energy or information between different parts of a nonautonomous, nonconservative system. This differs from the interpretation of the Lebesgue--Stieltjes model of \cite{fra2}, where this macrosystem was regarded as conservative. Because of the greater level of control over all the details of the algebra, the present interpretation supersedes the one in Ref.\ \cite{fra2}. In multifractional spacetimes, varying the scale is equivalent to varying $\a$, so any relative change of probability between two scales $\ell$ and $\ell'$ will be reflected in a change of magnitude in the dissipation effect.


\section{Quantization of the free theory and free propagator}\label{propa}

We shall consider the free case, with potential $V=\frac12 m^2 \phi^2$. Then, the momentum algebra is satisfied, and $P^\mu$ can be interpreted as momentum and $-P^\mu P_\mu$ as the mass operator. Nonetheless, as a quantum operator, $\hat P^\mu $ cannot be associated with the derivative operator $\p^\mu$ but, rather, to $-\rmi   \cD^\mu$; see Eq.\ \Eq{phatdef}.

Since the functions \Eq{E}  are eigenstates of $\hat P^\mu =-\rmi \cD^\mu$, it is natural to decompose 
the field $\phi(x)$ according to \Eq{fmt}. Then, the fractional equation of motion $\cD_\mu \cD^\mu \phi (x)  -m^2 \phi(x)=0$ is mapped into an algebraic equation which is indistinguishable from the corresponding equation in usual nonfractional scalar theory, $(p^2+m^2)\tilde\phi(p)=0$, whose obvious solutions are 
\ba
\label{phitilp}
\tilde\phi(p) &=& \de (p^2+m^2) \phi (p)\nonumber \\
&=& \frac{1}{2\omega({\bf p})} \{\phi (\omega, {\bf p}) \, \de[p^0-\omega({\bf p})]\nonumber\\
&&\qquad\qquad + \phi (-\omega, {\bf p})\, \de[p^0 + \omega({\bf p})]\}\, .
\ea
Here $\omega({\bf p}) := \sqrt{{\bf p}^2+m^2}$, and $\phi(p) = \phi(p^0,{\bf p})$ is an arbitrary function evaluated on the support $p^2+m^2=0$. The inverse momentum transform \Eq{fmt} gives the field decomposition.

The canonical conjugate momentum is $\pi_\phi (x) = \cD_t \phi(x)$, so that, 
from Eqs.\ \Eq{pshi} and \Eq{E}, we have
\bs\label{vphipf} \ba
\label{vphif}
\phi (x)  &=& \frac{1}{\sqrt{v_0(t)}}\int \frac{\rmd \vr({\bf p})}{\sqrt{2\omega({\bf p})}}\left[a^\dagger ({\bf p}) \bE_v^* ({\bf p} ,x)\right.\nonumber\\
&&\qquad\left.\vphantom{a^\dagger}+a({\bf p}) \bE_v ({\bf p} , x)  \right]_{p^0=\omega({\bf p})},\\
\label{vphip}
\pi_\phi (x)  &=& \frac{i}{\sqrt{v_0(t)}}\int \rmd \vr({\bf p}) \sqrt{\frac{\omega({\bf p})}{2}} \left[a^\dagger ({\bf p}) \bE_v^* ({\bf p} ,x)\right.\nonumber\\
&&\qquad \left.\vphantom{a^\dagger}-a({\bf p}) \bE_v ({\bf p} , x) \right]_{p^0=\omega({\bf p})},
\ea
\es
where
\be
\label{aalpha}
a({\bf p}) =  \frac{\phi(\omega,{\bf p})}{\sqrt{4\pi \omega({\bf p}) v({\bf p})}}
\ee
and 
\be
\label{Ebold}
\bE_v ({\bf p} , x)= \left. \frac{\rme^{i p\cdot x}}{(2\pi)^{\frac{D-1}2}\, \sqrt{v({\bf x})v({\bf p})}}\right|_{p^0=\omega({\bf p})}.
\ee
According to our conventions on fractional integrations when only space variables are involved, $\bE_v ({\bf p} , x)$ is defined and normalized in such a way that 
\bs\label{Eboldnorm}\ba
&&\left. \int d\vr ({\bf p}) \, \bE_v^* ({\bf p},x)\, \bE_v ({\bf p},y)\right|_{x^0=y^0}\nonumber\\
&&\qquad=\de_v ({\bf x},{\bf y})= \frac{\delta({\bf x-y})}{\sqrt{v({\bf x})v({\bf y})}},\\
&&\left. \int d\vr ({\bf x}) \, \bE_v^* ({\bf p},x)\, \bE_v ({\bf k},x)\right|_{x^0=0}\nonumber\\
&&\qquad=\de_v ({\bf p},{\bf k}) = \frac{\delta({\bf p-k})}{\sqrt{v({\bf p})v({\bf k})}}.
\ea\es

In fractional spaces the correspondence principle relating the commutator of two quantum operators in the Heisenberg picture and the fractional canonical bracket of the corresponding classical observables is [see Eq.\ \Eq{poipoi}]
\be
\label{cp}
v_0(t)[A(t),B(t)] = \rmi \hbar \{ A(t), B(t)\}_v\,,
\ee
leading to the canonical equal time commutators ($\hbar=1$)
\be
 \label{heifrac}
 \left[ \phi(t,{\bf x}), \pi_\phi (t, {\bf y})\right]= \frac{i}{v_0(t)} \, \de_v ({\bf x,y})\,,
 \ee
with the other commutators vanishing.
This implies the algebra 
\bs\label{algea}\ba
\left[a ({\bf p}),a^\dagger ({\bf p'})\right]&=& \de_v ({\bf p,p'})\,,\\
\left[a^\dagger ({\bf p}),a^\dagger ({\bf p'})\right]&=& 0\,,\\
\left[a^{ } ({\bf p}),a^{ } ({\bf p'})\right]&=& 0\,.
\ea\es
Accordingly, the pair $(a({\bf p}),\,a^\dagger ({\bf p}))$ has the same interpretation of annihilation and creation operators in fractional spaces as in usual Minkowski spacetimes. Introducing the fractional vacuum state $|0 \rangle_v$ as the only eigenstate of the operator $a ({\bf p})$ with a vanishing eigenvalue, it is possible to create a Fock space through repeated action of $a^\dagger$ operators on the vacuum state, $a^\dagger({\bf p}_1) \cdots a^\dagger({\bf p}_n)|0\rangle_v$. Such states are eigenstates of the (normal ordered) momentum operator
\ba
 P^\nu &=& \int \rmd \vr({\bf x}): T^{0\nu}\!:\nonumber\\
 &=& \frac{1}{v_0(t)}\int \rmd\vr({\bf p}) \,\left.  p^\nu  a^\dagger ( {\bf p})  a( {\bf p}) \right|_{p^0= \omega ({\bf p})}\,,
\ea
with eigenvalues $p^\nu = \left(\sum_{i=1}^n \omega ({\bf p}_i) \, , \sum_{i=1}^n {\bf p}_i\right)$. The normalization of the one-particle state $ |{\bf p}\rangle_v = a^\dagger ({\bf p}) | 0\rangle_v$ is  $\langle {\bf p}|{\bf k}\rangle_v = \de_v ({\bf p, k})$.

The presence of the factors $1/\sqrt{v_0(t)}$ in Eqs.\ \Eq{vphif} and \Eq{vphip} is important. In the definition of $\phi$, this factor is necessary to render  $\sqrt{v(x)} \, \phi$ a free scalar field in Minkowski spacetime [see Eq.\Eq{pshi}]; the remaining $\sqrt{v({\bf x})}$ factor simplifies the corresponding weight hidden in the definition of $\bE_v({\bf p},x)$. Concerning $\pi_\phi$, the same factor makes the definition of conjugate momentum as weighted  derivative, $\pi_\phi (x) =\cD_t \phi (x) = [v(t)]^{-1/2} \p_t [\sqrt{v(t)} \, \phi(x)]$, consistent with the integer-picture commutators (see the Appendix, Sec.\ \ref{ippro}, for details). 

Finally, the Feynman two-point function is
\ba
G(x,y) &=& \rmi \langle 0 \left| \cT \left[ \phi(x) \phi (y) \right] \right|0\rangle_v\nonumber\\
&=& \int \rmd \vr (k) \,  \frac{\bE_v(k,x) \bE_v^*(k,y) }{k^2+ m^2 -\rmi \epsilon}=\frac{\bar G(x-y)}{\sqrt{v(x)v(y)}}\,,\nonumber\\\label{prophi}
\ea
where $\cT$ denotes the chronologically ordered product, integration is over the $D$-dimensional momentum space, $-\rmi \epsilon$ in the integrand enforces the causal prescription, and $\bar G$ is the ordinary propagator in integer spacetime. The $\bE_v(k,x)$ functions in \Eq{prophi} are those in Eq.\ \Eq{E}, which are eigenfunctions of the operator $-i \cD_\mu$ with eigenvalue $k_\mu$. Consequently, from \Eq{delfra} it follows that \Eq{prophi} is a solution of the (fractional) Green equation
\be
(-\cD_\mu \cD^\mu+m^2) G(x,y) = \de_v (x,y)\,.
\ee

Equation \Eq{prophi} states that the propagator in position space is the usual one times a measure prefactor. The pole structure of Eq.\ \Eq{prophi} is identical to the standard case, since the momentum-space measure is reabsorbed by the normalization of the $\bE_v(k,x)$. Therefore, we have the usual mass poles irrespective of the form of the measure, i.e., both in the fractional case \Eq{amea} for any $\a$ and in the multifractional case \Eq{muf2}. This means that the concept of particle field with mass $m$ is meaningful in multifractional spacetimes and valid \emph{at any scale}, throughout the whole dimensional flow. Contrary to other approaches such as asymptotic safety \cite{LaR5} and the Lebesgue--Stieltjes toy model of Refs.\ \cite{fra1,fra2}, the propagator in position space is not logarithmic at the UV critical point, where the dimension of spacetime is 2. On the other hand, in Lifshitz-type theories, where higher-order spatial derivatives are present and dominate at different regimes, the $k^0$ (and hence pole) structure of the propagator remains the same at all scales \cite{Hor2,AH,Ans1,Vis09,IRS}, as in multifractional models (see \cite{fra7} for a comparison with asymptotic safety and Ho\v{r}ava--Lifshitz spacetimes).

Notice also that Eq.\ \Eq{prophi} can be regularized and then expressed as the integral of a probability density function, as in the integer case (e.g., \cite{Riv07}). In Euclidean signature, the kernel
\be
G_l(k):= \frac{\rme^{-l(k^2+ m^2)}}{k^2+ m^2}=\int_l^{+\infty}\rmd \s\,\rme^{-\s(k^2+ m^2)}
\ee
provides the regularized correlation function
\ba
G_l(x,y)&=&\int \rmd \vr (k) \,  G_l(k)\,\bE_v(k,x) \bE_v^*(k,y)\nonumber\\
&=&\int_l^{+\infty}\rmd \s\,\rme^{-\s m^2}u(x,y,\s)\,,
\ea
where $u(x,y,\s)=[v(x)v(y)]^{-1/2}(4\pi\s)^{-D/2}\exp[-|x-y|^2/(4\s)]$. One recognizes $u$ as the solution of the diffusion equation with nonanomalous diffusion time describing the ordinary Brownian motion of a point particle from point $x$ to point $y$ \cite{frc4}. It is the ordinary Gaussian distribution of a Wiener process times a measure prefactor.


\section{Discussion}\label{disc}

After studying the symmetry structure of a classical scalar field theory in multifractional Minkowski spacetime, we have quantized the field in terms of creation and annihilation operators and found the Feynman propagator. Although ordinary Poincar\'e invariance is broken, there exists a Poincar\'e algebra of generators of symmetries for the free theory, which makes it possible to define the concept of mass as in ordinary spacetimes. Factorizability of the coordinate dependence of the measure [Eq.\ \Eq{genf}] is crucial for consistency. A facultative tool one can employ is the formal equivalence, at the level of the action in position space, between systems in fractional spacetimes and a class of nonautonomous systems with ordinary measure $\rmd^Dx$, where a certain explicit coordinate dependence is present in the Lagrangian. At least for a scalar field theory, multifractal spacetimes provide a physical and geometric interpretation of a quantum field theory with spacetime-dependent couplings.

Multifractional measures can be defined as factorizable, thus entering the picture automatically and providing an important extension of the work done here and previously. All the results are then applied not only at a given scale [$v(x)=v_\a(x)$, Eq.\ \Eq{amea}], but at any scale as well [Eq.\ \Eq{muf2}]. As we pointed out in the previous section, the mass poles in the propagator persist all the way through the deep UV ($\a\to 2/D$, Hausdorff dimension $\dh=2$) where geometry, however, is nonconformal. If we started with a Lagrangian with a fractional kinetic term of order $2\g$, the eigenvalue $-k^2$ in the propagator would be replaced by $-k^{2\g}:= |k^0|^{2\g}-\sum_{i=1}^{D-1} |k^i|^{2\g}$ \cite{frc4,AIP}. This would give rise to a branch cut and a continuous spectrum of quasiparticle modes, thus rendering the interpretation of the quantum theory far less traditional than the present one with a second-order kinetic term. 

The study of quantum interactions and of the renormalization properties of fractional scalar field theory will be the next natural step.


\begin{acknowledgments}
The work of G.C.\ is under a Ram\'on y Cajal contract.

\end{acknowledgments}


\appendix*

\section{Integer picture}

All the results of the paper can be recast in the scalar (or integer) picture after the field redefinition \Eq{pshi}.
Before doing so in this Appendix, we stress that one cannot abandon the knowledge of the nontrivial position measure after the trick \Eq{pshi}. Volume integrals, inner products and many other ingredients of the theory are consistently defined before any field redefinition. The free theory is only \emph{formally} identical to the integer one, because as soon as one looks into details in the model one recognizes that there are differences even in the absence of interactions; some of these differences are discussed below Eq.\ \Eq{vr}, others in Ref.\ \cite{frc5}. Here we recall only that the definition of the action symmetries and of the position coordinate scaling (and, hence, of a nontrivial momentum space) fixes once and for all the geometry and scale identifications. Thus, even if one can get rid of the measure weights at the level of the free action in the present model of fractional spacetimes,\footnote{In other fractional models the Laplacian is either non-self-adjoint or a genuine nonlocal fractional operator (e.g., Ref.\ \cite{AIP}), and there is no field redefinition that can reabsorb the measure weight and the nontrivial differential structure.} one cannot forfeit the geometric structure of the model \emph{tout court}. Neither the field redefinition \Eq{pshi} nor any other change of variable can affect the geometry of the system.

\subsection{Equations of motion}

One imposes $\de S=0$ in Eq.\ \Eq{Spsi} under a functional variation of the field $\vp\to\vp+\de\vp$ and ends up with the usual equation
\be \label{eomvp}
\frac{\p\bar\cL}{\p\vp}-\p_\mu\frac{\p\bar\cL}{\p(\p_\mu\vp)}=0\,,
\ee
corresponding, for a power-law potential $V(\vp)\propto \vp^n$, to
\be\label{eompsi} 
\p_\mu\p^\mu\vp-[v(x)]^{1-\frac{n}2} V_{,\vp}(\vp)=0\,.
\ee
This agrees, via \Eq{pshi}, with Eq.\ \Eq{eomph}.

\subsection{Energy-momentum tensor}

Here we recover the (non)conservation law of the energy-momentum tensor, Eq.\ \Eq{contphi2}, for a power-law potential $V(\vp)=\lambda \vp^n$. Under an infinitesimal coordinate and field transformation  $\de x^\mu=x'^\mu - x^\mu$, $\de_0 \vp (x) = \vp'(x')-\vp(x)$, the variation of the action can be obtained along the same lines of the standard Noether theorem:
\ba  
\de_0 S &=&\int\left[\de(\rmd^Dx)\bar \cL+\rmd^Dx\,\de_0 \bar \cL\right]\nonumber\\
& \stackrel{\Eq{dede}} {=}\ &\int\rmd^Dx\left[\bar \cL\p_\mu\de x^\mu +(\de \bar \cL+\p_\mu\bar \cL\de x^\mu)\right]\nonumber\\
&=&\int\rmd^Dx\left[\p_\mu(\bar \cL\de x^\mu)+ \frac{\p\bar \cL}{\p \vp} \de\vp +\frac{\p\bar \cL}{\p \p_\mu \vp} \de\p_\mu \vp \right]\nonumber\\
&=&\int\rmd^Dx\left[\p_\mu(\bar \cL\de x^\mu)+ \left(\frac{\p \bar \cL}{\p\vp}\de\vp-\p^\mu \vp \p_\mu\de\vp\right)\right]\nonumber\\
&\ \stackrel{\Eq{eompsi}}{=}\ &\int\rmd^Dx\,  \p_\mu \left[\bar \cL\de x^\mu-\p^\mu \vp \de\vp
\right]\nonumber\\
&\stackrel{\Eq{dede}} {=}\ &\int\rmd^Dx\,\p_\mu\left[\left( \de^\mu_\nu\, \bar \cL + \p^\mu \vp \p_\nu \vp \right)\de x^\nu
- \p^\mu\vp \de_0\vp\right]\,.\nonumber\\
\ea
The integrand in the right-hand side is (minus) the  divergence of the standard Noether current which, in the case of translations $\de x^\mu = -\epsilon^\mu $ and $\de_0 \vp (x)=0$, is just the energy-momentum tensor for the field $\vp$:
\be\label{tilT}
\bar T^\mu {}_{\nu} := v T_{\mu\nu}=\de^\mu{}_\nu\, \bar\cL+\p^\mu\vp\p_\nu\vp\,.
\ee

The nonautonomous potential breaks translation invariance, leading to 
\ba
\de_0 S &=& \int \rmd^D x'\, \bar \cL'(x')-\int \rmd^D x\, \bar\cL(x)\nonumber\\
          &=& -\lambda \int \rmd^D x [\vp(x)]^n \left\{ [v(x-\e)]^{1-n/2} - [v(x)]^{1-n/2} \right\}\nonumber\\
          &=& \lambda \left(1-\frac{n}{2} \right) \int \rmd^D x\, \e^\nu \p_\nu  v\, v^{-n/2} [\vp(x)]^n\, .
        \ea
Consequently, the energy-momentum tensor is not conserved except in the $n=2$ case, and the standard continuity equation is replaced by 
\be\label{cont}
\p_\mu \bar T^\mu{}_\nu = \left(\frac{n}2-1\right)\la\,\p_\nu v\,v^{-n/2}\vp^n =: s_\nu(x,\vp)\,.
\ee
Integrating Eq.\ \Eq{cont} in space, we get the integer-picture analog of \Eq{contphi3},
\be\label{cont2}
\dot{\bar P}^\nu=\int \rmd{\bf x}\,s^\nu(x,\vp)\,,
\ee
where the momentum $D$-vector in the integer picture is defined as
\be
\bar P^\nu := \int \rmd{\bf x}\, \bar T^{0\nu}=v_0(t)\, P^\nu\,.
\ee
Let $\bar H :=\bar P^0$ be the Hamiltonian of the system. From the usual Hamilton evolution equation $\dot{\bar A} =\p_t \bar A+\{\bar A,\bar H\}$, one infers, as in the main text, that the time-space translation algebra does not close. In fact, $\{\bar P_i,\bar H\}=\dot{\bar P}_i\neq 0$ unless $n=2$. Here one uses the Hamilton equations
\be\label{hamps}
\dot\vp =\{\vp,\bar H\}\,,\qquad \dot\pi_\vp =\{\pi_\vp,\bar H\}\,,
\ee
where a dot denotes the total derivative $\rmd_t=\rmd/\rmd t$ and $\pi_\vp=\dot\vp=\sqrt{v}\pi_\phi$ is the momentum conjugate to $\vp$.

\subsection{Free propagator}\label{ippro}

We can take advantage of the fact that in the free case it is always possible to redefine the field in such a way as to reabsorb measure factors completely. Consequently, its quantization is straightforward. Equation \Eq{phitilp} is the same in both fractional and integer pictures. From that, one has
\ba
\vp (x) &=& \frac{1}{(2\pi)^{\frac{D-1}2}}\int \rmd {\bf p}\,  \frac{1}{\sqrt{2\omega({\bf p})}}\left[\bar a^\dagger ({\bf p}) \rme^{-\rmi p x}\right.\nonumber\\
&&\qquad\left. + \bar a ({\bf p})  \rme^{\rmi p x} \right]_{p^0=\omega({\bf p})}\label{vphik}
\ea
and $\pi_\vp (x)=\dot \vp (x)$, where $\bar a({\bf p}) =  \phi(\omega,{\bf p}) /{\sqrt{4\pi \omega({\bf p})}}$. Quantization in the scalar picture is identical to that of scalar fields in Minkowski spacetime. The correspondence principle 
$[\,\cdot\,,\cdot\,]= \rmi \, \{\,\cdot\,,\cdot\,\}$ induces the usual Heisenberg algebra for the equal-time canonical commutation relations:
\bs\label{algealpha}\ba
&&\left[\bar a ({\bf p}),\bar a^\dagger ({\bf p'})\right]= \delta ({\bf p-p'})\,,\\
&&\left[\bar a^\dagger ({\bf p}),\bar a^\dagger ({\bf p'})\right]= 0,\qquad
\left[\bar a^{ } ({\bf p}),\bar a^{ } ({\bf p'})\right]= 0.
\ea\es
Under the setting $\bar a({\bf p}) = a({\bf p}) \, \sqrt{v({\bf p})}$, the holomorphic algebra \Eq{algealpha} is mapped into its natural generalization to fractional spaces \Eq{algea}, i.e., the same algebra with delta functions replaced by fractional ones. Consistently, upon multiplication by $v_0(t) \sqrt{v({\bf x}) v({\bf y})}$, Eq.\ \Eq{heifrac} reproduces the correct  commutator $[\vp(t,{ \bf x}),\pi_\vp (t,{\bf y})]=i\, \delta({\bf x-y})$ of the scalar picture in Minkowski spacetime.

The Fock space is constructed as usual. With the conventions of \Eq{vphik}, the normalization  of the one-particle state $|{\bf p} \rangle = \bar a^\dagger({\bf p}) |0\rangle$ is  $\langle {\bf p'} | {\bf p }\rangle = \delta ({\bf p-p'})$. 

Finally, the difference between fields in the scalar and fractional pictures is a $c$-number, so that there is no need to modify the chronological ordering operator $\cT$, as
\be
\cT[\phi(x) \phi(y)] = \frac{1}{\sqrt{v(x)v(y)}} \cT[\vp(x)\vp(y)]\,.
\ee
Consequently, it is easy to verify that the free Feynman propagator in integer picture is
\ba
\bar G (x-y)&=& \rmi \langle 0 \left| \cT \left[ \vp(x) \vp (y) \right] \right|0\rangle\nonumber\\
&=&\int\frac{\rmd^Dk}{(2\pi)^D}\, \frac{\rme^{\rmi k\cdot (x-y)}}{k^2+ m^2 -\rmi \epsilon}\,.\label{provp}
\ea
Equation \Eq{provp} defines the free Feynman propagator, and it clearly satisfies the Green equation $(-\B + m^2)\bar G(x-y)=\delta(x-y)$. 



\begin{thebibliography}{99}

%
\bibitem{Wil73} K.G.\ Wilson, \tia{Quantum field-theory models in less than 4 dimensions} \doin{10.1103/PhysRevD.7.2911}{Phys.\ Rev.\ D}{7}{2911}{1973}.
\bibitem{tHo93} G.\ 't Hooft, \tia{Dimensional reduction in quantum gravity} in \emph{Salamfestschrift}, edited by A.\ Ali, J.\ Ellis, and S.\ Randjbar-Daemi (World Scientific, Singapore, 1993) [\oarX{gr-qc/9310026}].
\bibitem{Car09} S.\ Carlip, \tia{Spontaneous dimensional reduction in short-distance quantum gravity?} 
  \doin{10.1063/1.3284402}{AIP Conf.\ Proc.}{1196}{72}{2009} [\arX{0909.3329}].
\bibitem{fra1}  G.\ Calcagni, \tia{Fractal universe and quantum gravity} \doin{10.1103/PhysRevLett.104.251301}{Phys.\ Rev.\ Lett.}{104}{251301}{2010} [\arX{0912.3142}].
\bibitem{Car10} S.\ Carlip, \tia{The small scale structure of spacetime} in {\it Foundations of Space and Time}, edited by G.\ Ellis, J.\ Murugan, and A.\ Weltman (Cambridge University Press, Cambridge, England, 2012) [\arX{1009.1136}].

\bibitem{AJL4}  J.\ Ambj{\o}rn, J.\ Jurkiewicz, and R.\ Loll, \tia{Spectral dimension of the universe}
\doin{10.1103/PhysRevLett.95.171301}{Phys.\ Rev.\ Lett.}{95}{171301}{2005} [\oarX{hep-th/0505113}].
\bibitem{BeH}   D.\ Benedetti and J.\ Henson, \tia{Spectral geometry as a probe of quantum spacetime} \doin{10.1103/PhysRevD.80.124036}{Phys.\ Rev.\ D}{80}{124036}{2009} [\arX{0911.0401}].
\bibitem{GWZ1}  G.\ Giasemidis, J.F.\ Wheater, and S.\ Zohren, \tia{Dynamical dimensional reduction in toy models of $4D$ causal quantum gravity} \doin{10.1103/PhysRevD.86.081503}{Phys.\ Rev.\ D}{86}{081503(R)}{2012}
[\arX{1202.2710}].
\bibitem{GWZ2}  G.\ Giasemidis, J.F.\ Wheater, and S.\ Zohren, \tia{Multigraph models for causal quantum gravity and scale dependent spectral dimension} \doin{10.1088/1751-8113/45/35/355001}{J.\ Phys.\ A}{45}{355001}{2012} [\arX{1202.6322}].
\bibitem{LaR5}  O.\ Lauscher and M.\ Reuter, \tia{Fractal spacetime structure in asymptotically safe gravity} \doij{10.1088/1126-6708/2005/10/050}{J.\ High Energy Phys.} {10}{050}{2005} [\oarX{hep-th/0508202}].
\bibitem{ReS11} M.~Reuter and F.~Saueressig, \tia{Fractal space-times under the microscope: a renormalization group view on Monte Carlo data} \doij{10.1007/JHEP12(2011)012}{J.\ High Energy Phys.}{12}{012}{2011} [\arX{1110.5224}].
\bibitem{Mod08} L.\ Modesto, \tia{Fractal structure of loop quantum gravity}
\doin{10.1088/0264-9381/26/24/242002}{Classical Quantum Gravity}{26}{242002}{2009} [\arX{0812.2214}].
\bibitem{CaM}   F.\ Caravelli and L.\ Modesto, \tia{Fractal dimension in 3d spin-foams} \arX{0905.2170}.
\bibitem{MPM}   E.\ Magliaro, C.\ Perini, and L.\ Modesto, \tia{Fractal space-time from spin-foams} \arX{0911.0437}.
\bibitem{COT}   G.\ Calcagni, D.\ Oriti, and J.\ Th\"urigen, \tia{Laplacians on discrete and quantum geometries} \arX{1208.0354}; and (work in progress).
\bibitem{Hor2}  P.\ Ho\v{r}ava, \tia{Quantum gravity at a Lifshitz point}
 \doin{10.1103/PhysRevD.79.084008}{Phys.\ Rev.\ D}{79}{084008}{2009} [\arX{0901.3775}].
\bibitem{Hor3}  P.\ Ho\v{r}ava, \tia{Spectral dimension of the universe in quantum gravity at a Lifshitz point} \doin{10.1103/PhysRevLett.102.161301}{Phys.\ Rev.\ Lett.}{102}{161301}{2009} [\arX{0902.3657}].
\bibitem{SVW1}  T.P.\ Sotiriou, M.\ Visser, and S.\ Weinfurtner, \tia{Spectral dimension as a probe of the ultraviolet continuum regime of causal dynamical triangulations} \doin{10.1103/PhysRevLett.107.131303}{Phys.\ Rev.\ Lett.}{107}{131303}{2011} [\arX{1105.5646}].
\bibitem{Con06} A.\ Connes, \tia{Noncommutative geometry and the standard model with neutrino mixing}
 \doij{10.1088/1126-6708/2006/11/081}{J.\ High Energy Phys.}{11}{081}{2006} [\oarX{hep-th/0608226}].
\bibitem{CCM}   A.H.\ Chamseddine, A.\ Connes, and M.\ Marcolli, \ndoin{http://intlpress.com/site/pub/pages/journals/items/atmp/content/vols/0011/0006/00024717/index.html}{Adv.\ Theor.\ Math.\ Phys.}{11}{991}{2007} [\oarX{hep-th/0610241}].
\bibitem{Ben08} D.\ Benedetti, \tia{Fractal properties of quantum spacetime} \doin{10.1103/PhysRevLett.102.111303}{Phys.\ Rev.\ Lett.}{102}{111303}{2009} [\arX{0811.1396}].
\bibitem{AA}    E.\ Alesci and M.\ Arzano, \tia{Anomalous dimension in semiclassical gravity} \doin{10.1016/j.physletb.2011.12.026}{Phys.\ Lett.\ B}{707}{272}{2012} [\arX{1108.1507}].
\bibitem{ACOS}  M.\ Arzano, G.\ Calcagni, D.\ Oriti, and M.\ Scalisi, \tia{Fractional
and noncommutative spacetimes} \doin{10.1103/PhysRevD.84.125002}{Phys.\ Rev.\ D}{84}{125002}{2011} [\arX{1107.5308}].
\bibitem{Mod11} L.\ Modesto, \tia{Super-renormalizable quantum gravity} \doin{10.1103/PhysRevD.86.044005}{Phys.\ Rev.\ D}{86}{044005}{2012} [\arX{1107.2403}].
\bibitem{BGKM}  T.\ Biswas, E.\ Gerwick, T.\ Koivisto, and A.\ Mazumdar, \doin{10.1103/PhysRevLett.108.031101}{Phys.\ Rev.\ Lett.}{108}{031101}{2012} [\arX{1110.5249}].
\bibitem{AMM}   S.\ Alexander, A.\ Marcian\`o, and L.\ Modesto, \tia{The hidden quantum groups symmetry of super-renormalizable gravity} \doin{10.1103/PhysRevD.85.124030}{Phys.\ Rev.\ D}{85}{124030}{2012} [\arX{1202.1824}].
\bibitem{Mod12} L.\ Modesto, \tia{Super-renormalizable multidimensional quantum gravity} \arX{1202.3151}.
\bibitem{MoN}   L.\ Modesto and P.\ Nicolini, \doin{10.1103/PhysRevD.81.104040}{Phys.\ Rev.\ D}{81}{104040}{2010} [\arX{0912.0220}].
\bibitem{SSN}   E.\ Spallucci, A.\ Smailagic, and P.\ Nicolini, \doin{10.1103/PhysRevD.73.084004}{Phys.\ Rev.\ D}{73}{084004}{2006} [\oarX{hep-th/0604094}].
\bibitem{Mur12} J.R.\ Mureika, \doin{10.1016/j.physletb.2012.08.029}{Phys.\ Lett.\ B}{716}{171}{2012} [\arX{1204.3619}].
\bibitem{MuN}  J.\ Mureika and P.\ Nicolini, \arX{1206.4696}.

\bibitem{Sti77} F.H.\ Stillinger, \tia{Axiomatic basis for spaces with noninteger dimension}  
 \doin{10.1063/1.523395}{J.\ Math.\ Phys.\ (N.Y.)}{18}{1224}{1977}.
\bibitem{Svo87} K.\ Svozil, \tia{Quantum field theory on fractal space-time}
 \doin{10.1088/0305-4470/20/12/033}{J.\ Phys.\ A}{20}{3861}{1987}.
\bibitem{Ey89a} G.\ Eyink, \tia{Quantum field-theory models on fractal spacetime. I: Introduction and overview}
 \doin{10.1007/BF01228344}{Commun.\ Math.\ Phys.}{125}{613}{1989}.
\bibitem{Ey89b} G.\ Eyink, \tia{Quantum field-theory models on fractal spacetime. II: Hierarchical propagators}
 \doin{10.1007/BF02124332}{Commun.\ Math.\ Phys.}{126}{85}{1989}.
\bibitem{Gol08} E.\ Goldfain, \tia{Fractional dynamics and the TeV regime of field theory}
 \doin{10.1016/j.cnsns.2006.06.001}{Commun.\ Nonlinear Sci.\ Numer.\ Simul.}{13}{666}{2008}.
\bibitem{fra2}  G.\ Calcagni, \tia{Quantum field theory, gravity and cosmology in a fractal universe} \doij{10.1007/JHEP03(2010)120}{J.\ High Energy Phys.}{03}{120}{2010} [\arX{1001.0571}].
\bibitem{fra3}  G.\ Calcagni, \tia{Gravity on a multifractal}
\doin{10.1016/j.physletb.2011.01.063}{Phys.\ Lett.\ B}{697}{251}{2011} [\arX{1012.1244}]. 

\bibitem{frc1}  G\ Calcagni, \tia{Geometry of fractional spaces} \ndoin{http://intlpress.com/site/pub/pages/journals/items/atmp/content/vols/0016/0002/00024226/index.html}{Adv.\ Theor.\ Math.\ Phys.}{16}{549}{2012} [\arX{1106.5787}].
\bibitem{frc2}  G.\ Calcagni, \tia{Geometry and field theory in multi-fractional spacetime}
 \doij{10.1007/JHEP01(2012)065}{J.\ High Energy Phys.}{01}{065}{2012}  [\arX{1107.5041}].
\bibitem{frc3}  G.\ Calcagni and G.\ Nardelli, \tia{Momentum transforms and Laplacians in fractional spaces} Adv.\ Theor.\ Math.\ Phys. \textbf{16} (to be published) [\arX{1202.5383}].
\bibitem{frc4}  G.\ Calcagni, \tia{Diffusion in multiscale spacetimes} \doin{10.1103/PhysRevE.87.012123}{Phys.\ Rev.\ E}{87}{012123}{2013} [\arX{1205.5046}].
\bibitem{frc5}  G.\ Calcagni, G.\ Nardelli, and M.\ Scalisi, \tia{Quantum mechanics in fractional and other anomalous spacetimes} \doin{10.1063/1.4757647}{J.\ Math.\ Phys.\ (N.Y.)}{53}{102110}{2012} [\arX{1207.4473}].
\bibitem{fra7}  G.\ Calcagni, \tia{Multi-fractional spacetimes, asymptotic safety and Ho\v{r}ava--Lifshitz gravity} \arX{1209.4376}.
\bibitem{fra4}  G.\ Calcagni, \tia{Discrete to continuum transition in multifractal spacetimes} \doin{10.1103/PhysRevD.84.061501}{Phys.\ Rev.\ D}{84}{061501(R)}{2011} [\arX{1106.0295}].
\bibitem{fra6}  G.\ Calcagni, \tia{Diffusion in quantum geometry} \doin{10.1103/PhysRevD.86.044021}{Phys.\ Rev.\ D}{86}{044021}{2012} [\arX{1204.2550}].
\bibitem{AIP}   G.\ Calcagni, \tia{Introduction to multifractional spacetimes} \doin{10.1063/1.4756961}{AIP Conf.\ Proc.}{1483}{31}{2012} [\arX{1209.1110}].

%
\bibitem{RYS}   F.-Y.\ Ren, Z.-G.\ Yu, and F.\ Su, \tia{Fractional integral associated to the self-similar set or the generalized self-similar set and its physical interpretation} \doin{10.1016/0375-9601(96)00418-5}{Phys.\ Lett.\ A}{219}{59}{1996}.
\bibitem{YRZ}   Z.-G.\ Yu, F.-Y.\ Ren, and J.\ Zhou, \tia{Fractional integral associated to generalized cookie-cutter set and its physical interpretation} \doin{10.1088/0305-4470/30/15/036}{J.\ Phys.\ A}{30}{5569}{1997}.
\bibitem{RYZLM} F.-Y.\ Ren, Z.-G.\ Yu, J.\ Zhou, A.\ Le M\'ehaut\'e, and R.R.\ Nigmatullin, \tia{The relationship between the fractional integral and the fractal structure of a memory set} \doin{10.1016/S0378-4371(97)00353-1}{Physica (Amsterdam)}{246A}{419}{1997}.
\bibitem{Yu99}  Z.-G.\ Yu, \tia{Flux and memory measure on net fractals} \doin{10.1016/S0375-9601(99)00316-3}{Phys.\ Lett.\ A}{257}{221}{1999}.
\bibitem{QL} W.-Y.\ Qiu and J.\ L\"u, \tia{Fractional integrals and fractal structure of memory sets} \doin{10.1016/S0375-9601(00)00448-5}{Phys.\ Lett.\ A}{272}{353}{2000}.
\bibitem{RQLW} F.-Y.\ Ren, W.-Y.\ Qiu, J.-R.\ Liang, and X.-T.\ Wang, \tia{Determination of memory function and flux on fractals} \doin{10.1016/S0375-9601(01)00544-8}{Phys.\ Lett.\ A}{288}{79}{2001}.
\bibitem{RLWQ} F.-Y.\ Ren, J.-R.\ Liang, X.-T.\ Wang, and W.-Y.\ Qiu, \tia{Integrals and derivatives on net fractals} \doin{10.1016/S0960-0779(02)00211-4}{Chaos Solitons Fractals}{16}{107}{2003}.
\bibitem{NLM}   R.R.\ Nigmatullin and A.\ Le M\'ehaut\'e, \tia{Is there geometrical/physical meaning of the fractional integral with complex exponent?} \doin{10.1016/j.jnoncrysol.2005.05.035}{J.\ Non-Cryst.\ Solids}{351}{2888}{2005}.

%
\bibitem{HX}    H.-J.\ He and Z.-Z.\ Xianyu, \tia{Spontaneous spacetime reduction and unitary weak boson scattering at the LHC} \doin{10.1016/j.physletb.2013.01.06}{Phys.\ Lett.\ B}{720}{142}{2013} [\arX{1112.1028}].
\bibitem{Tar3}  V.E.\ Tarasov, \tia{Continuous medium model for fractal media} \doin{10.1016/j.physleta.2005.01.024}{Phys.\ Lett.\ A}{336}{167}{2005} [\oarX{cond-mat/0506137}].
\bibitem{Tar4b} V.E.\ Tarasov, \tia{Possible experimental test of continuous medium model for fractal media}
 \doin{10.1016/j.physleta.2005.05.022}{Phys.\ Lett.\ A}{341}{467}{2005}  [\oarX{physics/0602121}].
\bibitem{Tar7}  V.E.\ Tarasov, \tia{Wave equation for fractal solid string} \doin{10.1142/S0217984905008712}{Mod.\ Phys.\ Lett.\ B}{19}{721}{2005} [\oarX{physics/0605006}].
\bibitem{Fal03} K.\ Falconer, \textit{Fractal Geometry} (Wiley, New York, 2003).
\bibitem{Pod02} I.\ Podlubny, \tia{Geometric and physical interpretation of fractional integration and fractional differentiation} \ndoin{http://www.diogenes.bg/fcaa/}{Fract.\ Calc.\ Appl.\ Anal.}{5}{367}{2002} [\oarX{math.CA/0110241}].
\bibitem{tat95} F.B.\ Tatom, \tia{The relationship between fractional calculus and fractals} \doin{10.1142/S0218348X95000175}{Fractals}{03}{217}{1995}.
\bibitem{Sok12} I.M.\ Sokolov, \tia{Models of anomalous diffusion in crowded environments}
  \doin{10.1039/C2SM25701G}{Soft Matter}{8}{9043}{2012}.
\bibitem{MeK}   R.~Metzler and J.~Klafter, \tia{The random walk's guide to anomalous diffusion: a fractional dynamics approach} \doin{10.1016/S0370-1573(00)00070-3}{Phys.\ Rep.}{339}{1}{2000}.
\bibitem{Zas3}  G.M.~Zaslavsky, \tia{Chaos, fractional kinetics, and anomalous transport}
 \doin{10.1016/S0370-1573(02)00331-9}{Phys.\ Rep.}{371}{461}{2002}.
\bibitem{MeK2}  E.\ Metzler and J.\ Klafter, \tia{The restaurant at the end of the random walk: recent developments in the description of anomalous transport by fractional dynamics} \doin{10.1088/0305-4470/37/31/R01}{J.\ Phys.\ A}{37}{R161}{2004}.
\bibitem{Tar12} V.E.\ Tarasov, \tia{Fractional vector calculus and fractional Maxwell's equations} \doin{10.1016/j.aop.2008.04.005}{Ann.\ Phys.\ (Amsterdam)}{323}{2756}{2008} [\arX{0907.2363}].
\bibitem{CSN1}  K.\ Cottrill-Shepherd and M.\ Naber, \tia{Fractional differential forms} \doin{10.1063/1.1364688}{J.\ Math.\ Phys.\ (N.Y.)}{42}{2203}{2001} [\oarX{math-ph/0301013}].

%
\bibitem{Hut81} J.E.\ Hutchinson, \tia{Fractals and self similarity} \doin{10.1512/iumj.1981.30.30055}{Indiana Univ.\ Math.\ J.}{30}{713}{1981}.
\bibitem{Har01} D.\ Harte, \textit{Multifractals: Theory and Applications} (Chapman and Hall/CRC, Boca Raton, 2001).
\bibitem{Sor98} D.\ Sornette, \tia{Discrete scale invariance and complex dimensions} \doin{10.1016/S0370-1573(97)00076-8}{Phys.\ Rep.}{297}{239}{1998} [\oarX{cond-mat/9707012}].
\bibitem{LvF}   M.L.\ Lapidus and M.\ van Frankenhuysen, \textit{Fractal Geometry, Complex Dimensions and Zeta Functions} (Springer, New York, 2006).
\bibitem{Akk1}  E.\ Akkermans, G.V.\ Dunne, and A.\ Teplyaev, \tia{Physical consequences of complex dimensions of fractals} \doin{10.1209/0295-5075/88/40007}{Europhys.\ Lett.}{88}{40007}{2009} [\arX{0903.3681}].
\bibitem{Akk2}  E.\ Akkermans, G.V.\ Dunne, and A.\ Teplyaev, \tia{Thermodynamics of photons on fractals} \doin{10.1103/PhysRevLett.105.230407}{Phys.\ Rev.\ Lett.}{105}{230407}{2010} [\arX{1010.1148}].
\bibitem{Akk12} E.\ Akkermans, \tia{Statistical mechanics and quantum fields on fractals} \arX{1210.6763}.

\bibitem{Pel91} Y.\ Peleg, \tia{Change of dimensions in canonical pure gravity via nonunitarity}
 \doin{10.1142/S0217732391003018}{Mod.\ Phys.\ Lett.\ A}{06}{2569}{1991}.
\bibitem{Rie1}  F.E.\ Riewe, \tia{Nonconservative Lagrangian and Hamiltonian mechanics} \doin{10.1103/PhysRevE.53.1890}{Phys.\ Rev.\ E}{53}{1890}{1996}.
\bibitem{Rie2}  F.E.\ Riewe, \tia{Mechanics with fractional derivatives} \doin{10.1103/PhysRevE.55.3581}{Phys.\ Rev.\ E}{55}{3581}{1997}.
\bibitem{El05a} R.A.\ El-Nabulsi, \tia{A fractional action-like variational approach of some classical, quantum and geometrical dynamics} Int.\ J.\ Appl.\ Math.\ {\bf 17}, 299 (2005).
\bibitem{El05b} R.A.\ El-Nabulsi, \tia{A fractional approach to nonconservative Lagrangian dynamical systems} \ndoin{http://fizika.hfd.hr/fizika_a/av05/a14p289.htm}{Fiz.\ A}{14}{289}{2005}.

%
\bibitem{AH}    D.\ Anselmi and M.\ Halat, \tia{Renormalization of Lorentz violating theories}  
 \doin{10.1103/PhysRevD.76.125011}{Phys.\ Rev.\ D}{76}{125011}{2007} [\arX{0707.2480}].
\bibitem{Ans1}  D.\ Anselmi, \tia{Weighted scale invariant quantum field theories}
 \doij{10.1088/1126-6708/2008/02/051}{J.\ High Energy Phys.}{02}{051}{2008} [\arX{0801.1216}].
\bibitem{Vis09} M.\ Visser, \tia{Lorentz symmetry breaking as a quantum field theory regulator}  
 \doin{10.1103/PhysRevD.80.025011}{Phys.\ Rev.\ D}{80}{025011}{2009} [\arX{0902.0590}].
\bibitem{IRS}   R.\ Iengo, J.G.\ Russo, and M.\ Serone, \tia{Renormalization group in Lifshitz-type theories} \doij{10.1088/1126-6708/2009/11/020}{J.\ High Energy Phys.}{11}{020}{2009} [\arX{0906.3477}].
\bibitem{Riv07} V.\ Rivasseau, \tia{Non-commutative renormalization} \arX{0705.0705}.
\end{thebibliography}
\end{document}